\documentclass[11pt,onecolumn,prc,showpacs,preprintnumber,amsmath,
floatfix,amssymb]{revtex4}
\usepackage{epsfig}
\usepackage{graphicx}
\usepackage{dcolumn}
\usepackage{bm}
\def\r{\mbox{{\bf  r}}}
\def\p{\mbox{\boldmath $p$}}
\def\q{\mbox{\boldmath $q$}}
\def\k{\mbox{\boldmath $k$}}
\def\t{\mbox{\boldmath $t$}}
\allowdisplaybreaks[4]

\begin{document}
\title{Analysis of quasi-elastic neutrino charged-current scattering off 
$^{16}$O and neutrino energy reconstruction}
\author{A.~V.~Butkevich}
\affiliation{ Institute for Nuclear Research,
Russian Academy of Sciences,
60th October Anniversary Prosp. 7A,
Moscow 117312, Russia}
\date{\today}
\begin{abstract}

The charged-current quasi-elastic scattering of muon neutrino on the oxygen
target is analyzed for neutrino energy up to 2.5 GeV using
the Relativistic Distorted-Wave Impulse Approximation (RDWIA). 
The inclusive cross sections $d^2\sigma/dQ^2$, calculated within the RDWIA, 
are lower than  the Relativistic Fermi Gas Model (RFGM) results in the range of
the square of four-momentum transfer $Q^2\leq$0.2 (GeV/c)$^2$. We have 
also studied the nuclear-model dependence of the neutrino energy 
reconstruction accuracy using the charged-current quasi-elastic events
 with no detector effects and background. We found that for one-track
 events the accuracy is nuclear-model dependent for neutrino energy up 
to 2.5 GeV.

\end{abstract}

\maketitle

\section{Introduction}

The field of neutrino oscillations is rapidly developing. The goals of 
the current and planed set of accelerator-based neutrino experiments
~\cite{K2KA,MiniA,Mino,Miner,T2K,OPERA,NOVA} are the precision measurements of
observed neutrino mass splitting and the detailed study of the neutrino mixing
matrix. The data of these experiments will greatly extend the statistics due to
extremely intense neutrino beamline.

To study the neutrino oscillation effects on the terrestrial distance
scale, the neutrino beams cover the energy range from a
few hundred MeV to several GeV. In this energy range, the dominant 
contribution to the neutrino-nucleus cross section comes from the 
charged-current (CC) quasi-elastic (QE) reactions and resonance 
production processes. 
The cross section data in this energy range are rather 
scarce and were taken on the targets, which are not used in the neutrino 
oscillation experiments (i.e.  water, iron, lead or plastic).

In this situation, the statistical uncertainties should be negligible as 
compared to systematic errors in the incident neutrino flux, neutrino 
interaction model and the detector effects on the neutrino events selection and
neutrino energy reconstruction. Apparently, these uncertainties produce
systematic errors in the extraction of oscillation parameters.

To evaluate the neutrino mass squared difference in the muon neutrino
disappearance experiments, the probability of $\nu_{\mu}$ disappearance 
versus neutrino energy is measured. Because the CCQE interaction represents 
a two-particle scattering process, it forms a good signal sample, and the 
neutrino energy may be estimated using the kinematics of this reaction. 
There are two ways of measuring  the neutrino energy: either kinematic or 
calorimetric reconstruction. In detectors with the energy threshold for 
proton detection $\varepsilon^p_{th} \geq1$ GeV (Cherenkov detectors) 
the muon neutrino CCQE interactions will  produce the one-track events, i.e. 
only muons are detected in the final states. The kinematic reconstruction is 
applied for these events. Assuming the target nucleon to be at rest inside the
nucleus, the correlation between the incident neutrino energy and a
reconstructed muon momentum and scattering angle is used in this method. 

 In the fine-grained detectors the two-track CCQE events are detected, and the 
calorimetric reconstruction can be applied, if the particle identification of
the second track and the resolution for proton momentum are reliable. In this 
case the visible neutrino energy is simply a sum of the reconstructed muon 
energy and kinematic proton energy. In this paper we consider the  procedures 
for neutrino energy reconstruction, which are based on the kinematics of the 
CCQE interaction. 

 In general, the detector efficiency and energy response are highly dependent 
on the
type of interaction: QE or non-QE (the resonance and deep inelastic 
scattering). 
The Monte Carlo (MC) simulation of the detector response to neutrino 
interactions is used for tuning the values of cuts for separation of 
the QE and non-QE (nQE) events and for estimating the efficiency of detecting 
the CCQE events after all cuts. To model the scattering from a nuclei, 
the most part of an event generator~\cite{Zel} is based on the 
RFGM~\cite{Moniz}, in which the nucleus is described as a system of quasi-free 
nuclei with a flat nucleon momentum distribution up to the same Fermi momentum 
$p_F$ and nuclear binding energy $\epsilon_b$. But this model does not take 
into account the nuclear shall structure, the final state interaction (FSI) 
between the outgoing nucleon and residual nucleus and the presence of 
short-range nucleon-nucleon (NN) correlations, leading to appearance of a 
high-momentum and high-energy component in the nucleon momentum-energy 
distribution in the target.

The comparison with the high-precision electron scattering data 
has shown ~\cite{But1} that the accuracy of the RFGM prediction becomes 
poor at low $Q^2$, where the nuclear effects are largest, and this
model fails~\cite{But2} in application to exclusive cross sections. The modern
quasi-elastic neutrino scattering  data (the CCQE event distribution as 
a function of $Q^2$)~\cite{MiniA,Gran,Kator} also reveal the inadequacies 
in the present neutrino cross section simulation. The data/MC disagreement 
shows the data deficit in the low $Q^2$ region ($Q^2 \leq 0.2$ GeV$^2$) and 
the data excess in the high $Q^2$ region. The disagreement at low $Q^2$ 
would eventually result in the data/MC disagreement in the reconstructed 
neutrino energy.  

The Relativistic Distorted-Wave Impulse Approximation, which takes into account
the nuclear shall structure and FSI effects, was developed for description of 
electron-nucleus scattering, and it was successfully tested against the 
data~\cite{Fissum}. The RDWIA approach was also applied to neutrino-nucleus 
($\nu A$) interactions  for calculating the exclusive and inclusive QE cross 
sections~\cite{But2,Meucci,Maieron,Martin}. In Ref.~\cite{But2} the FSI 
effects on the inclusive cross section in the presence of the NN-correlations 
were estimated. 

    The aim of this work is twofold. First, we compute the RDWIA CCQE cross 
section versus $Q^2$ for muon neutrino scattering off oxygen. Second, we show 
the nuclear-model dependence of the efficiency of two-track CCQE events 
selection. We also estimate systematic uncertainties in the reconstructed 
neutrino energy within the RDWIA and RFGM taking into account the nucleon 
momentum distribution in the target, i.e. the nucleon Fermi motion effect.

The outline of this paper is as follows. In Sec. II we present briefly the
formalism for CCQE scattering process and the RDWIA model. The nuclear-model
dependence of cuts, which are applied for CCQE events selection, as well as the
 neutrino energy reconstruction methods are discussed in Sec. III. The results 
of numerical calculations are presented in Sec. IV. Our conclusions are 
summarized in Sec. V. In Appendix A we present the equation for neutrino 
energy, and in Appendix B the expressions for the moments of the reconstructed 
neutrino energy distribution are given. 

\section{Formalism of quasi-elastic scattering and models}

We consider the electron and neutrino charged-current QE exclusive
\begin{equation}\label{qe:excl}
\nu(k_i) + A(p_A)  \rightarrow l(k_f) + N(p_x) + B(p_B),      
\end{equation}
and inclusive
\begin{equation}\label{qe:incl}
\nu(k_i) + A(p_A)  \rightarrow l(k_f) + X                      
\end{equation}
scattering-off nuclei in the one-W-boson exchange approximation. Here $l$
represents the scattered lepton (electron or muon), $k_i=(\varepsilon_i,\k_i)$ 
and $k_f=(\varepsilon_f,\k_f)$ are the initial and final lepton 
momenta, $p_A=(\varepsilon_A,\p_A)$, and $p_B=(\varepsilon_B,\p_B)$ are 
the initial and final target momenta, $p_x=(\varepsilon_x,\p_x)$ is the 
ejectile nucleon momentum, $q=(\omega,\q)$ is the momentum transfer carried by 
the virtual W-boson, and $Q^2=-q^2=\q^2-\omega^2$ is the W-boson virtuality. 

\subsection{ CCQE neutrino-nucleus cross sections}

In the laboratory frame the differential cross section for the exclusive
(anti)neutrino CCQE reaction, in which only a single discrete state or narrow
resonance of the target is excited, can be written as
\begin{equation}
\label{cs5:cc}
\frac{d^5\sigma}{d\varepsilon_f d\Omega_f d\Omega_x} = R
\frac{\vert\p_x\vert\varepsilon_x}{(2\pi)^5}\frac{\vert\k_f\vert}       
{\varepsilon_i} \frac{G^2\cos^2\theta_c}{2} L_{\mu \nu}W^{\mu \nu},
\end{equation}
 where $\Omega_f$ is the solid angle for the lepton momentum, $\Omega_x$ is the
 solid angle for the ejectile nucleon momentum, 
$G \simeq$ 1.16639 $\times 10^{-11}$ MeV$^{-2}$ is
the Fermi constant, $\theta_C$ is the Cabbibo angle
($\cos \theta_C \approx$ 0.9749), $L_{\mu \nu}$ and $W^{\mu \nu}$ are,
respectively, the lepton and weak CC nuclear tensors. The recoil factor $R$ 
is given by
\begin{equation}\label{Rec}
R =\int d\varepsilon_x \delta(\varepsilon_x + \varepsilon_B - \omega -m_A)=
{\bigg\vert 1- \frac{\varepsilon_x}{\varepsilon_B}
\frac{\p_x\cdot \p_B}{\p_x\cdot \p_x}\bigg\vert}^{-1},                    
\end{equation}
 and $\varepsilon_x$ is the solution to the equation
\begin{equation}\label{eps}
\varepsilon_x+\varepsilon_B-m_A-\omega=0,                                 
\end{equation}
where $\varepsilon_B=\sqrt{m^2_B+\p^2_B}$, $~\p_B=\q-\p_x$, $~\p_x=
\sqrt{\varepsilon^2_x-m^2}$, and $m_A$, $m_B$, and $m$ are masses of the 
target, recoil nucleus and nucleon, respectively. 
The missing momentum $p_m$ and missing energy $\varepsilon_m$ are defined by 
\begin{subequations}
\begin{align}
\label{p_m}
\p_m & = \p_x-\q
\\
\label{eps_m}
\varepsilon_m & = m + m_B - m_A                                           
\end{align}
\end{subequations}
The lepton tensor can be written as a sum of symmetric $L^{\mu\nu}_S$ and
antisymmetric $L^{\mu\nu}_A$ tensors
\begin{subequations}
\begin{align}
\label{Lmunu}
L^{\mu\nu} &= L^{\mu\nu}_S + L^{\mu\nu}_A
\\
\label{Lmunu:S}
L^{\mu\nu}_S &= 2\left(k^{\mu}_i k^{\nu}_f + k^{\nu}_i k^{\mu}_f -
g^{\mu\nu}k_ik_f\right)\\                                                
\label{Lmunu:A}
L^{\mu\nu}_A &= h2i\epsilon^{\mu\nu\alpha\beta}(k_i)_{\alpha}(k_f)_{\beta},
\end{align}
\end{subequations}
where $h$ is $+1$ for a positive lepton helicity, and $-1$ for a negative 
lepton helicity, $\epsilon^{\mu \nu \alpha \beta}$ is the antisymmetric tensor
with $\epsilon^{0 1 2 3}=-\epsilon_{0 1 2 3}=1$. 
The weak CC hadronic tensors $W_{\mu \nu}$ are given by bilinear products 
of the transition matrix elements of the nuclear CC operator $J_{\mu}$ between 
the initial nucleus state $\vert A \rangle $ and the final state  
$\vert B_f \rangle$ as
\begin{eqnarray}
W_{\mu \nu } &=& \sum_f \langle B_f,p_x\vert                           
J_{\mu}\vert A\rangle \langle A\vert
J^{\dagger}_{\nu}\vert B_f,p_x\rangle,              
\label{W}
\end{eqnarray}
where the sum is taken over undetected states.

In the inclusive reactions (\ref{qe:incl}) only the outgoing lepton is
detected, and the differential cross sections can be written as
\begin{equation}
\frac{d^3\sigma}{d\varepsilon_f d\Omega_f } =
\frac{1}{(2\pi)^2}\frac{\vert\k_f\vert}
{\varepsilon_i} \frac{G^2\cos^2\theta_c}{2} L_{\mu \nu}                  
\mathcal {W}^{\mu \nu },
\end{equation}
where $\mathcal{W}^{\mu \nu}$ is an inclusive hadronic tensor.
In the reference frame, in which the {\it z} axis is parallel to the
momentum transfer $\q=\k_i - \k_f$ and the {\it y} axis is
parallel to $\k_i \times \k_f$, the exclusive neutrino
scattering cross sections take the forms
\begin{widetext}
\begin{align}\label{cs:excl}
\frac{d^5\sigma}{d\varepsilon_f d\Omega_f d\Omega_x} &=
\frac{\vert\p_x\vert\varepsilon_x}{(2\pi)^5}G^2\cos^2\theta_c             
\varepsilon_f \vert \k_f \vert R \big \{ v_0R_0 + v_TR_T
 + v_{TT}R_{TT}\cos 2\phi + v_{zz}R_{zz}
\notag \\
& +(v_{xz}R_{xz} - v_{0x}R_{0x})\cos\phi  
-v_{0z}R_{0z} + h\big[v_{yz}(R^{\prime}_{yz}\sin\phi + R_{yz}\cos\phi)
\notag \\
& - v_{0y}(R^{\prime}_{0y}\sin\phi + R_{0y}\cos\phi) - v_{xy}R_{xy}\big]\big\},
\end{align}
\end{widetext}
where $\theta$, $\varphi$ are lepton scattering angles, and $\theta_{x}$,
$\phi$ are outgoing nucleon angles, $v_i$ are the neutrino coupling
coefficients, and $R_i$ are independent response functions~\cite{But2}, which
depend on the variables $Q^2$, $\omega$, $|\p_x|$, and $\theta_x$. 
Similarly, the inclusive lepton scattering cross sections are reduced to
\begin{equation}
\frac{d^3\sigma}{d\varepsilon_f d\Omega_f} =
\frac{G^2\cos^2\theta_c}{(2\pi)^2} \varepsilon_f
\vert \k_f \vert\big ( v_0R_0 + v_TR_T+ v_{zz}R_{zz} -v_{0z}R_{0z}- 
hv_{xy}R_{xy}\big)
\end{equation}
where the response functions now depend on $Q^2$ and $\omega$ only 
~\cite{But2}. 

We describe the lepton-nucleon scattering in the Impulse 
Approximation (IA), in which only one nucleon of the target is involved in 
the reaction, and  the nuclear current is written as a sum of single-nucleon 
currents. Then, the nuclear matrix element in Eq.(\ref{W}) takes the form
\begin{eqnarray}\label{Eq12}
\langle p,B\vert J^{\mu}\vert A\rangle &=& \int d^3r~ \exp(i\t\cdot\r)
\overline{\Psi}^{(-)}(\p,\r)
\Gamma^{\mu}\Phi(\r),                                                     
\end{eqnarray}
where $\Gamma^{\mu}$ is the vertex function, $\t=\varepsilon_B\q/W$ is the
recoil-corrected momentum transfer, $W=\sqrt{(m_A + \omega)^2 - \q^2}$ is the
invariant mass, $\Phi$ and $\Psi^{(-)}$ are relativistic bound-state and
outgoing wave functions.

The single-nucleon charged current has $V{-}A$ structure $J^{\mu} = 
J^{\mu}_V + J^{\mu}_A$. For the free-nucleon vertex function 
$\Gamma^{\mu} = \Gamma^{\mu}_V + \Gamma^{\mu}_A$ we use the CC2 vector 
current vertex function ~\cite{Fore}
\begin{equation}
\Gamma^{\mu}_V = F_V(Q^2)\gamma^{\mu} + {i}\sigma^{\mu \nu}\frac{q_{\nu}}
{2m}F_M(Q^2)                                                            
\end{equation}
and the axial current vertex function
\begin{equation}
\Gamma^{\mu}_A = F_A(Q^2)\gamma^{\mu}\gamma_5 + _P(Q^2)q^{\mu}\gamma_5,  
\end{equation}
where $\sigma^{\mu \nu}=i[\gamma^{\mu}, \gamma^{\nu}]/2$.
The weak vector form factors $F_V$ and $F_M$ are related with corresponding
electromagnetic factors for proton $F^{(el)}_{i,p}$ and neutron 
$F^{(el)}_{i,n}$ by the hypothesis of conserved vector current (CVC)
\begin{equation}
F_i = F^{(el)}_{i,p} - F^{(el)}_{i,n},                                 
\end{equation}
where $F^{(el)}_V$ and $F^{(el)}_M$ are the Dirac and Pauli nucleon form 
factors. We use the approximation of~\cite{MMD} for these form factors. 
Because the bound nucleons are off shell, we employ the de Forest 
prescription for off-shell vertex~\cite{Fore} and the Coulomb gauge - 
for vector current $J_V$.

The axial $F_A$ and psevdoscalar $F_P$ form factors in the dipole
approximation are parameterized as
\begin{equation}
F_A(Q^2)=\frac{F_A(0)}{(1+Q^2/M_A^2)^2},\quad                          
F_P(Q^2)=\frac{2m F_A(Q^2)}{m_{\pi}^2+Q^2},
\end{equation}
where $F_A(0)=1.267$, $M_A\simeq 1.032$ GeV is the axial mass, and $m_{\pi}$ 
is the pion mass 

\subsection{ Models}

In the independent particle shell model the relativistic bound-state functions 
$\Phi$ 
 in Eq.(\ref{Eq12}) are obtained within the Hartree--Bogolioubov approximation 
in the $\sigma-\omega$ model ~\cite{Serot}.  The bound-state spinor takes the 
form
\begin{equation}\label{Eq17}
\Phi_{\kappa m}({\bf r}) = \left(
\begin{array}{r}
F_{\kappa}(r) {\cal Y}_{\kappa m}(\hat{r}) \\                           
i G_{-\kappa}(r) {\cal Y}_{-\kappa m}(\hat{r})
\end{array} \right),
\end{equation}
where
\begin{equation} 
{\cal Y}_{\kappa m}(\hat{r}) = \sum_{\nu,m_s}
\Big\langle \begin{array}{ll} \ell & \frac{1}{2} \\ \nu & m_s \end{array} 
\large | 
\begin{array}{l} j \\ m \end{array} \Big\rangle 
Y_{\ell\nu}(\hat{r})\chi_{m_s}
\end{equation}
is the spin spherical harmonic, and the orbital and total angular momenta are 
given, respectively, by
\begin{subequations}
\begin{eqnarray}
\ell &=& S_\kappa(\kappa + \frac{1}{2}) - \frac{1}{2} \\                  
j &=& S_\kappa \kappa - \frac{1}{2}
\end{eqnarray}
\end{subequations}
with $S_\kappa=sign{(\kappa)}$. The missing momentum distribution is then 
\begin{equation}
P(p_m) = \frac{S_{\alpha}}{2\pi^2} \left( 
|\tilde{F}_{\kappa}(p_m)|^2 + |\tilde{G}_{\kappa}(p_m)|^2 \right),       
\end{equation}
where
\begin{subequations}
\begin{eqnarray}
\label{Eq21}
\tilde{F}_{\kappa}(p) &=& \int dr \; r^2  j_{\ell}(p_m r) F_{\kappa}(r) \\
\tilde{G}_{-\kappa}(p) &=& \int dr \; r^2  j_{\ell^\prime}(p_m r)         
G_{-\kappa}(r),                                                         
\end{eqnarray}
\end{subequations}
and $j_{\ell}(x)$ is the Bessel function of order $\ell$, and
$\ell^{\prime}=2j-\ell$. In this work the bound-nucleon wave functions 
~\cite{NLSH} are used in the numerical analysis with the normalization factors 
$S_{\alpha}$ relative to full occupancy of $^{16}$O: $S(1p_{3/2})=0.66$,
~$S(1p_{1/2})=0.7$~\cite{Fissum} and $S(1s_{1/2})=1$. Note that the calculation
of the bound-nucleon wave function for $1p_{3/2}$-state includes the
incoherent contribution of the unresolved $2s_{1/2}d_{5/2}$ 
doublet~\cite{Leus}. 

In the RDWIA the ejectile wave function $\Psi$ in Eq.(\ref{Eq12}) is obtained 
following the direct Pauli reduction method~\cite{Udi,Hed}. It is well known 
that the Dirac spinor
\begin{eqnarray}
\Psi = 
\begin{pmatrix}                                                          
\Psi _+ \\ 
\Psi_-
\end{pmatrix}
\end{eqnarray}
can be written in terms of its positive energy component $\Psi _+$ as
\begin{eqnarray}
\Psi = 
\begin{pmatrix}  \Psi _+ \\
 \frac {\bm{\sigma} \cdot \p}{E+M+S-V} \Psi _+,                          
\end{pmatrix}
\end{eqnarray}
where $S=S(r)$ and $V=V(r)$ are the scalar and vector potentials for the
 nucleon with energy $E$. The upper component $\Psi _+$ can be related to the  
Schr\"odinger-like wave function $\xi $ by the Darwin factor $D(r)$, i.e.
\begin{eqnarray}
\Psi _+ = \sqrt {D(r)}\ \xi  \label{eq.scheq},
\end{eqnarray}                                                            
\begin{eqnarray}
D(r) = \frac {E+M+S(r)-V(r)}{E+M} \label{eq.Darwin}.                      
\end{eqnarray}
The two-component wave function $\xi $ is the solution of the Schr\"odinger 
equation containing equivalent central and spin-orbit potentials, which are 
functions of the scalar and vector potentials $S$ and $V$, and are energy 
dependent. We use the LEA program~\cite{LEA} for numerical calculation of the
distorted wave functions with EDAD1 SV relativistic optical potential
\cite{Coop}.

In the Plane-Wave Impulse Approximation (PWIA) the final state interaction 
between the outgoing nucleon and the residual nucleus is neglected, and the 
nonrelativistic PWIA exclusive cross section has a factorized form~\cite{Frul}
\begin{equation}
\frac{d^5\sigma}{d \varepsilon_f d\Omega_f d\Omega_x} = K\sigma_{ex}    
\mathcal{P}(E,\p) 
\end{equation}
where $K = Rp_x \varepsilon_x/(2\pi)^5$ is the phase-space factor and
$\sigma_{ex}$ is the half-off-shell cross section for neutrino scattering by 
a moving nucleon. The nuclear spectral function  $\mathcal{P}(E,\p)$ can be 
written as
\begin{equation}
\mathcal{P}(E,\p) = \sum_f \Big\vert\langle B_f\vert a(\p)\vert A\rangle\Big
\vert^2\delta(E-\varepsilon_m)                                          
\end{equation}
and the nucleon momentum distribution $P_{\beta}(\p)$ for the orbit 
$\beta$ is related to the upper component of the corresponding 
bound-state wave function (\ref{Eq21}) as
\begin{equation}
P_{\beta}(\p) = \frac{S_{\beta}}{2\pi^2} \Big\vert \tilde{F}_{\beta}(\p)  
\Big\vert^2.                                                            
\end{equation}

According to the JLab data \cite{Fissum}, the occupancy of the 
independent particle shell model orbitals of  $^{16}$O equals about 75\%, on 
the average. In this work we assume that the missing strength (25\%) can be 
attributed to the short-range NN-correlations in the ground state, leading to
appearance of high-momentum and high-energy nucleon distribution in the 
target. In order to estimate this effect in the inclusive cross sections, 
we consider the phenomenological model~\cite{Ciofi,Kul}
 where the high-momentum (HM) part of the spectral function is
determined by excited states with one or more nuclei in a continuum.

In our calculations of the inclusive cross sections only the real part of the
optical potential is included, because the complex potential produces 
absorptions of flux. Then, the contribution of the $1p$- and $1s$-states to the
inclusive cross section $\left(d^3\sigma/d\varepsilon_f d\Omega_f\right)
_{RDWIA}$ can be obtained by integrating the exclusive cross sections 
(\ref{cs:excl}) over $\Omega_x$. The effect of the FSI on the inclusive cross 
section can be evaluated using the ratio
\begin{equation} \label{Lambda-factor}
\Lambda(\varepsilon_f,\Omega_f) =
\bigg(\frac{d^3\sigma}{d\varepsilon_f d\Omega_f}\bigg)_{RDWIA}\bigg/      
\bigg(\frac{d^3\sigma}{d\varepsilon_f d\Omega_f}\bigg)_{PWIA},
\end{equation}
where $\left(d^3\sigma/d\varepsilon_f d\Omega_f\right)_{PWIA}$ is the 
result obtained in the PWIA. Then the total inclusive cross section can be 
written as
\begin{equation}
\frac{d^3\sigma}{d\varepsilon_f d\Omega_f} =
\bigg(\frac{d^3\sigma}{d\varepsilon_f d\Omega_f}\bigg)_{RDWIA} +
\Lambda(\varepsilon_f, \Omega_f)\bigg(\frac{d^3\sigma}                   
{d\varepsilon_f d\Omega_f}\bigg)_{HM},
\end{equation}
where $(d^3\sigma/d\varepsilon_f d\Omega_f)_{HM}$ is the high-momentum 
component contribution into the inclusive cross section~\cite{But2}.   

\section{Analysis of CCQE interaction and neutrino energy reconstruction }

\subsection{Differential cross sections $d\sigma/d\cos\theta$ and 
$d\sigma/dQ^2$}

The charged-current QE events distributions as a function of $Q^2$ or 
$\cos\theta$ were measured by K2K~\cite{Gran} and MiniBoone~\cite{MiniA,Kator}
experiments. High statistic data show a disagreement with the RFGM 
prediction. The data samples exhibit significant deficit in the region of low 
$Q^2\leq$0.2 GeV$^2$ and small muon scattering angles, which corresponds to 
forward-going muons. In Refs.~\cite{MiniA,Kator} it was shown that the data/MC 
disagreement is not due to mis-modeling of the incoming neutrino energy 
spectrum, but due to inaccuracy in the simulation of CCQE interactions. To 
tune the Fermi gas model to the low $Q^2$ region, an additional parameter was 
introduced, which reduced the phase volume of the nucleon Fermi gas at low 
momentum transfer. In the region of high $Q^2$ the data excess is observed, 
and the values of the axial vector mass $M_A$, obtained from a fit to the 
measured data, are higher, than the results of previous experiments.

We calculated the differential cross sections 
$d\sigma/d\cos\theta$ and $d^2\sigma/dQ^2$ for neutrino CCQE scattering off 
oxygen target in the RDWIA, PWIA and RFGM approaches. 

We note that in the case of (anti)neutrino scattering off free nucleon
CCQE the differential cross sections~\cite{Levi} 
$d\sigma^{\nu, \overline{\nu}}/dQ^2$ at $Q^2 \to $ 0 can be written as 
\begin{equation}
\frac{d\sigma^{\nu, \overline{\nu}}}{dQ^2} =\frac{G^2}{2\pi}\cos^2\theta_c
[F^2_V(0)+F^2_A(0)]                                                   
\end{equation}
and do not depend on the neutrino energy. The difference 
\begin{equation}
\frac{d\sigma^{\nu}}{dQ^2}-\frac{d\sigma^{\overline{\nu}}}{dQ^2} =
\frac{G^2}{\pi}\cos^2\theta_c\frac{Q^2}{m\varepsilon_i}
\left(1-\frac{Q^2}{4m\varepsilon_i}\right)(F_V+F_M)F_A                 
\end{equation}
is proportional to $F_A$ and decreases with neutrino energy. In the range of
$\varepsilon_i \sim$ 0.5 $\div$ 1 GeV it can be used for measuring the 
axial form factor $F_A$.

\subsection{Selection of charged-current QE two-track events}

At the first step, the CC candidate events are selected by requiring that at 
least one reconstructed track must be long and corresponding to a minimum 
ionizing particle with the momentum higher, than a few hundred MeVs. The 
background is
originated by neutral-current (NC) interactions producing a charged pion.

In the CC event candidates, the events with one or two reconstructed tracks, 
with vertex in the active target are selected like the CCQE events. No other 
tracks are allowed to be connected with this event vertex. The two-track 
events are divided into two samples: QE and nQE enriched samples. Depending on 
detector capabilities $dE/dx$, the information is applied to the second track 
for $\pi/p$ separation~\cite{SiBar}. Since the QE interaction is a 
two-particle scattering process, the measurement of the muon momentum and 
angle allows predicting the angle of a recoil proton (the second track) 
assuming the neutrino scattering off to occur with a nucleon at rest. If the 
measured second track agrees with this prediction 
within $\Delta \theta$, it represents likely the CCQE event. Using the MC 
simulation based on the Fermi gas model, the values of $\Delta \theta$ are 
chosen to give a reliable separation between the QE and nQE events.

To study the nuclear-model dependence of this cut, we consider the angle 
$\theta_{pq}$ between the direction of outgoing proton and momentum
transfer. For neutrino QE scattering off, the nucleon is at rest $\q=\p_x$ and
$\cos\theta_{pq}=$1. For scattering off bound nucleon with momentum $\p_m$, it 
follows from Eq.(\ref{p_m}), that   
\begin{equation}
\cos\theta_{pq} = \frac{\p^2_x + \q^2 - \p^2_m}{2|\p_x|\vert\q\vert}.     
\end{equation}
The maximum value of $\theta{pq}$ corresponds to scattering off nucleon with a 
maximum momentum $\p_{max}$, i.e. 
\begin{equation}
\label{cos}
\cos\theta^m_{pq} = \frac{\p^2_x + \q^2 -                        
\p^2_{max}}{2|\p_x|\vert\q\vert} 
\end{equation}
and $\cos\theta^m_{pq} \leq \cos\theta_{pq} \leq 1$.             

In the RFGM the recoil proton energy 
$\varepsilon_x=\sqrt{\p^2_m+m^2}-\epsilon_b+\omega$ and for  
$\vert\p_{max}\vert=p_F$ we have
\begin{equation}
\p^2_x = p^2_F + \tilde{\omega}^2 + 2\tilde{\omega}\sqrt{p^2_F+m^2},       
\end{equation}
where $\tilde{\omega}=\omega - \epsilon_b$. In the RDWIA the energy and
momentum of an outgoing nucleon can be written (see Eqs.(\ref{eps}),
(\ref{p_m})) as follow: 
\begin{subequations}
\begin{align}
\label{p_x}
\p_x & = \p_m+\q
\\                                                                        
\label{eps_x}
\varepsilon_x & = \omega + m_A - \varepsilon_B.                 
\end{align}
\end{subequations}
For the scattering off shell nucleon with a maximum momentum 
$\p_{max}$ the energy of recoil nuclei is 
\begin{equation}\label{eps_B}
\varepsilon_B = \sqrt{\p^2_{max} + m^2_B} \approx m_B + \p^2_{max}/2m_B,    
\end{equation}
where $m_B = m_A - m + \varepsilon_m$.                                         
In the numerical calculations we use $\vert\p_{max}\vert$=500 MeV/c and the 
mean missing energy $\langle\varepsilon_m\rangle$=27.1 MeV for the oxygen 
target. Using 
Eqs.(\ref{cos}), (\ref{p_x}), (\ref{eps_x}), and (\ref{eps_B}), we have
\begin{equation}
\cos\theta^m_{pq} = \frac{\bar{\omega}(2m+\bar{\omega})+(Q^2 - m^2)
-\p^2_{max}}{2\sqrt{\bar{\omega}(2m+\bar{\omega})(Q^2+m^2)}},               
\end{equation}
where 
$\bar{\omega}=\omega - \langle\varepsilon_m\rangle - \p^2_{max}/2m^{\star}_B$ 
and 
$m^{\star}_B = m_A - m + \langle\varepsilon_m\rangle$. It follows from 
(\ref{cos}), that in the RDWIA the phase volume in ($\cos\theta_{pq},Q^2$) 
coordinates is larger, than in the Fermi gas model, and this difference 
decreases with momentum transfer.

\subsection{Reconstruction of neutrino energy}

In the kinematic reconstruction the neutrino energy $\varepsilon_r$ is formed 
assuming 
the target nucleon to be at rest inside a nucleus
\begin{equation}\label{Rest}
\varepsilon_r=\frac{\varepsilon_f(m-\epsilon_b)-(\epsilon^2_b-2m\epsilon_b+
m^2_{\mu})/2}{(m-\epsilon_b)-\varepsilon_f+k_f\cos\theta}.                 
\end{equation}
This formula ignores the nucleon momentum distribution for the event 
reconstruction. 
Using Eq.(\ref{p_x}) and the energy balance in the RFGM  
\begin{equation}\label{eps_FG}
\varepsilon_i + \sqrt{\p^2_m + m^2} -\epsilon_b= \varepsilon_f+\varepsilon_x,
\end{equation}                                                             
or 
\begin{equation}\label{eps_RD}
\varepsilon_i + m_A = \varepsilon_x + \varepsilon_f + \varepsilon_B.      
\end{equation}
in the RDWIA for shell nucleon and 
\begin{equation}\label{eps_NN}
\varepsilon_i + \varepsilon_N = \varepsilon_f + \varepsilon_x.         
\end{equation}
for nucleons with energy $\varepsilon_N$ in the correlated NN-pair, we obtain
the second-order equation for the neutrino energy, which takes into account the
bound nucleon momentum and the energy distributions
\begin{equation}\label{Equ}
A\varepsilon^2_r - B\varepsilon_r + C= 0.                              
\end{equation}
The expressions for coefficients $A, B$, and $C$ are given in Appendix A
for the RFGM and RDWIA. The solution of (\ref{Equ})
\begin{equation}\label{Erec}
\varepsilon_r=\left(B+\sqrt{B^2 - 4AC}\right)/2A           
\end{equation}
is the reconstructed neutrino energy, which depends on the variables $|\p_m|$,
$\varepsilon_m$, and $\cos\tau=\p\cdot\q/|\p\cdot\q|$. So, the
distribution $\varepsilon_r(|\p_m|,\varepsilon_m,\cos\tau)$ corresponds to
measured values of ($k_f, \cos\theta$) and at $\varepsilon_m, \p_m \to$ 0
Eq.({\ref{Erec}}) has asymptotic form given by Eq.(\ref{Rest}).     

The $n$-th moment of $\varepsilon_r(k_f,\cos\theta,\p_m,\varepsilon_m)$
distribution versus $k_f$ and $\cos\theta$ can be written as
\begin{equation}\label{mom1}
\langle\varepsilon^n_r(k_f,\cos\theta)\rangle =
\int_{p_{min}}^{p_{max}}d\p \int_{\varepsilon_{min}}^{\varepsilon_{max}}
S(\p,\varepsilon)[\varepsilon_r(k_f,\cos\theta,\p,\varepsilon)]^n      
d\varepsilon,
\end{equation}
where $S(\p,\varepsilon)$ is the probability density function (p.d.f.) for the
nucleon momentum and energy, the target nucleon momentum and energy 
distribution being normalized with respect to the unit area. The mean of 
$\varepsilon_r(k_f,\cos\theta)$ and its variance $\sigma^2(\varepsilon_r)$ are 
defined by
\begin{subequations}
\begin{align}
\bar{\varepsilon}_r(k_f,\cos\theta) & = \langle\varepsilon_r
(k_f,\cos\theta)\rangle ,
\\                                                                       
\sigma^2(\varepsilon_r) & = 
\langle\varepsilon^2_r(k_f,\cos\theta)\rangle - 
\bar{\varepsilon}^2_r(k_f,\cos\theta)
\end{align}
\end{subequations}
In principle, the cut $
R=\sigma(\varepsilon_r)/\bar{\varepsilon}_r \leq \delta$ 
may be imposed (event by event) to select the events with well-reconstructed 
energy.

The accuracy of reconstructed energy $\varepsilon_r(\varepsilon_i)$ as a
function of $\varepsilon_i$ can be estimated using the moments of
$\varepsilon_r(k_f,\cos\theta)$ distribution 
\begin{equation}\label{mom}
\langle\varepsilon^n_r(\varepsilon_i)\rangle =
\int dk_f \int W(k_f,\cos\theta)                                        
[\varepsilon_r(k_f, \cos\theta)]^nd\cos\theta,
\end{equation}
where $W(k_f,\cos\theta)$ is the p.d.f. of the muon momentum and scattering
angle, i.e.            
\begin{equation}
W(k_f,\cos\theta) =\frac{1}{\sigma_{tot}(\varepsilon_i)}\frac{d^2\sigma}
{dk_f d\cos\theta},                                                     
\end{equation}
and
\begin{equation}
\sigma_{tot}(\varepsilon_i) =\int \frac{d^2\sigma}{dk_f d\cos\theta}dk_f 
d\cos\theta.                                                            
\end{equation}
Usually, to select the CC events, $k_f$ and $\cos\theta$ cuts are applied: 
$k_f\ge k_{cut}$ and $\cos\theta \geq (\cos\theta)_{cut}$. The lower limits of
integration in Eq.(\ref{mom}) are $(k_f)_{min}=k_{cut}$,  
$(\cos\theta)_{min}=(\cos\theta)_{cut}$ and 
$[\varepsilon_r(k_f,\cos\theta)]^n=\langle\varepsilon^n(k_f,\cos\theta)
\rangle$, if the nucleon Fermi motion effect is taken into account, or  
$\varepsilon_r(k_f,\cos\theta)$ is given by Eq.(\ref{Rest}), if this effect is
neglected. It's worth to emphasize here, that formula (\ref{Rest}) can not be
used for neutrino energy reconstruction at $\varepsilon_f \ge
(m-\epsilon_b)+k_f\cos\theta$ or $Q^2=Q^2_0 \ge 2m\varepsilon_i-m^2_{\mu}$, 
because the value of resulting $\varepsilon_r$ is negative in this region. 
In terms of 
energy transfer, it corresponds to the range $\omega_1 \leq \omega \leq
\omega_2$, where $\omega_2$ is the solution to the equation 
\begin{equation}\label{neg1}
Q^2_0 = 2\varepsilon_i(\varepsilon_f - k_f\cos\theta) - m^2_{\mu}.    
\end{equation}
In the RDWIA, $\omega_1$ is the value of $\omega$, at which
\begin{equation}\label{neg2}
Q^2_0 = \left[\vert\p_{max}\vert+\sqrt{\tilde{\omega}^2+
2m\tilde{\omega}}~\right]^2 -\omega^2                                 
\end{equation}
with $\tilde{\omega} = \omega -\langle\varepsilon_m\rangle- 
p^2_{max}/2m^{\star}_B$, and in the RFGM, $\omega_1$ is the solution to 
the equation
\begin{equation}
Q^2_0 = \left[p_F + \sqrt{p^2_F +2\epsilon_F\tilde{\omega} + 
\tilde{\omega}^2}~\right]^2-\omega^2,                                 
\end{equation}
where $\tilde{\omega} = \omega - \epsilon_b$ and $\epsilon^2_F = p^2_F +m^2$. 
The size of this range $\Delta\omega = \omega_2 - \omega_1$ is proportional to
$|\p_{max}|$ ($p_F$) and reduces with increasing $\cos\theta$.

The reconstructed neutrino energy 
$\bar{\varepsilon}_r=\langle\varepsilon_r\rangle$ is smeared with variance  
\begin{equation}
\sigma^2(\varepsilon_i)  = 
\langle\varepsilon^2_r(\varepsilon_i)\rangle -                        
\bar{\varepsilon}^2_r(\varepsilon_i)
\end{equation}
and biased with
\begin{equation}
\Delta(\varepsilon_i) = 
\varepsilon_i - \bar{\varepsilon}_r                                     
\end{equation}
Using this mean energy approach, we estimated the accuracy of the neutrino
energy reconstruction with and without the nucleon Fermi motion effect in the 
RDWIA and RFGM approaches. The expressions for the moments  
$\langle\varepsilon^n_r(k_f,\cos\theta)\rangle$ and 
$\langle\varepsilon^n_r(\varepsilon_i)\rangle$ are given in Appendix B.

In the calorimetric reconstruction  
$\varepsilon_r$ is formed as a sum of muon energy $\varepsilon_f$, kinematic 
proton energy $T_p$ and the mean missing energy $\langle\varepsilon_m\rangle$
\begin{equation}\label{calor}
\varepsilon(k_f,\cos\theta)=\varepsilon_f + T_p + \langle\varepsilon_m\rangle.
\end{equation}                                                        
The expressions for the moments of $\varepsilon_r(k_f,\cos\theta)$ 
distribution are given in Appendix B. 
However, the neutrino energy is underestimated in the kinematical and 
calorimetric reconstructions, when the event represents, in fact, the nQE 
event, but looks like the QE event. 
\begin{figure*}
  \begin{center}
    \includegraphics[height=16cm,width=16cm]{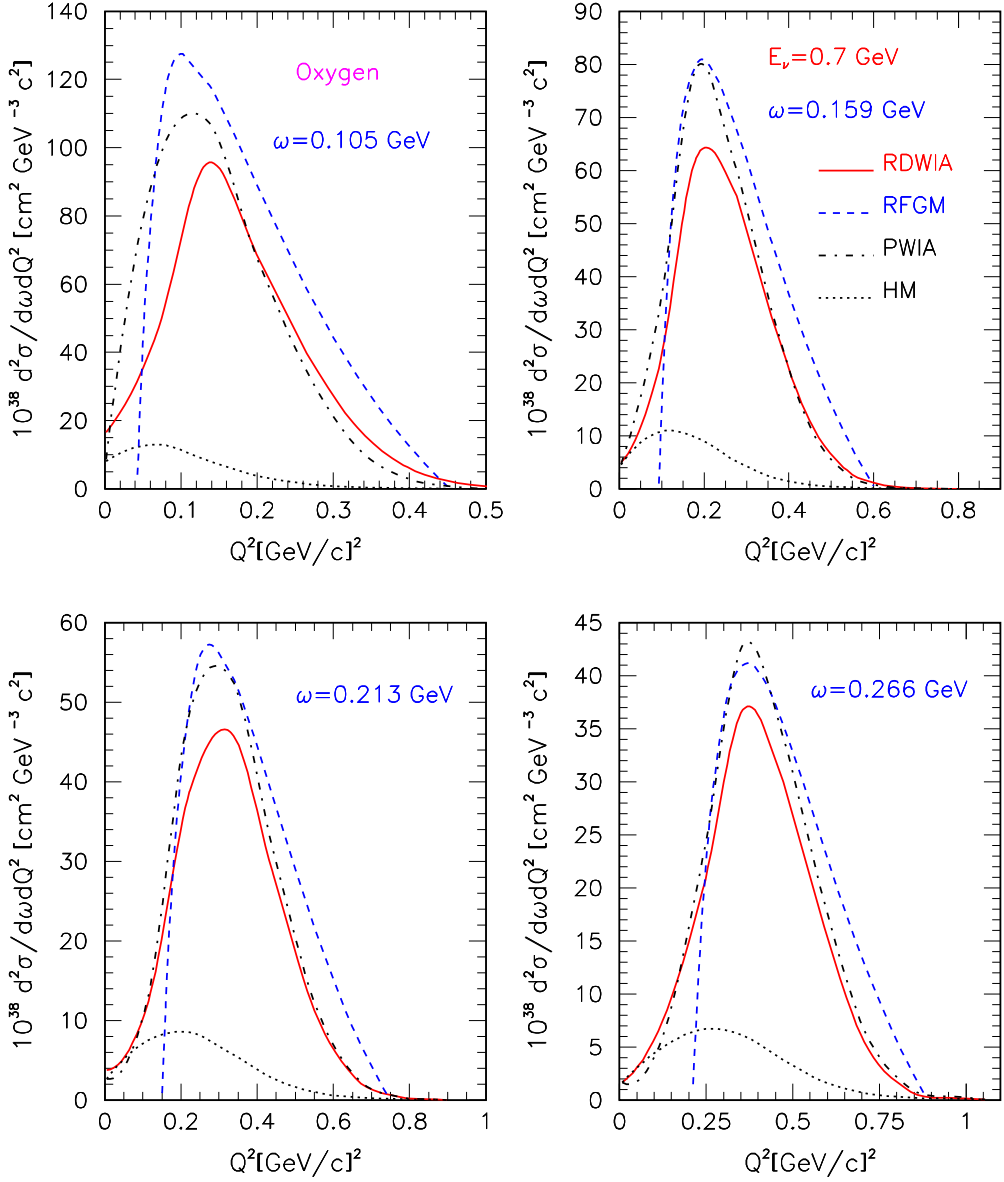}
\end{center}
\caption{(Color online) Inclusive cross section versus the four-momentum
  transfer $Q^2$ for neutrino scattering off $^{16}$O with energy
  $\varepsilon_{\nu}$=0.7 GeV and for four values of energy transfer: 
$\omega$=0.105, 0.159, 0.213 and 0.266 GeV. The solid line is the RDWIA
calculation, whereas the dashed and dash-dotted lines are, respectively, 
the RFGM and PWIA calculations. The dotted line is the high-momentum component 
contribution to the inclusive cross section.
}
\end{figure*}
\begin{figure*}
\begin{center}
\includegraphics[height=16cm,width=16cm]{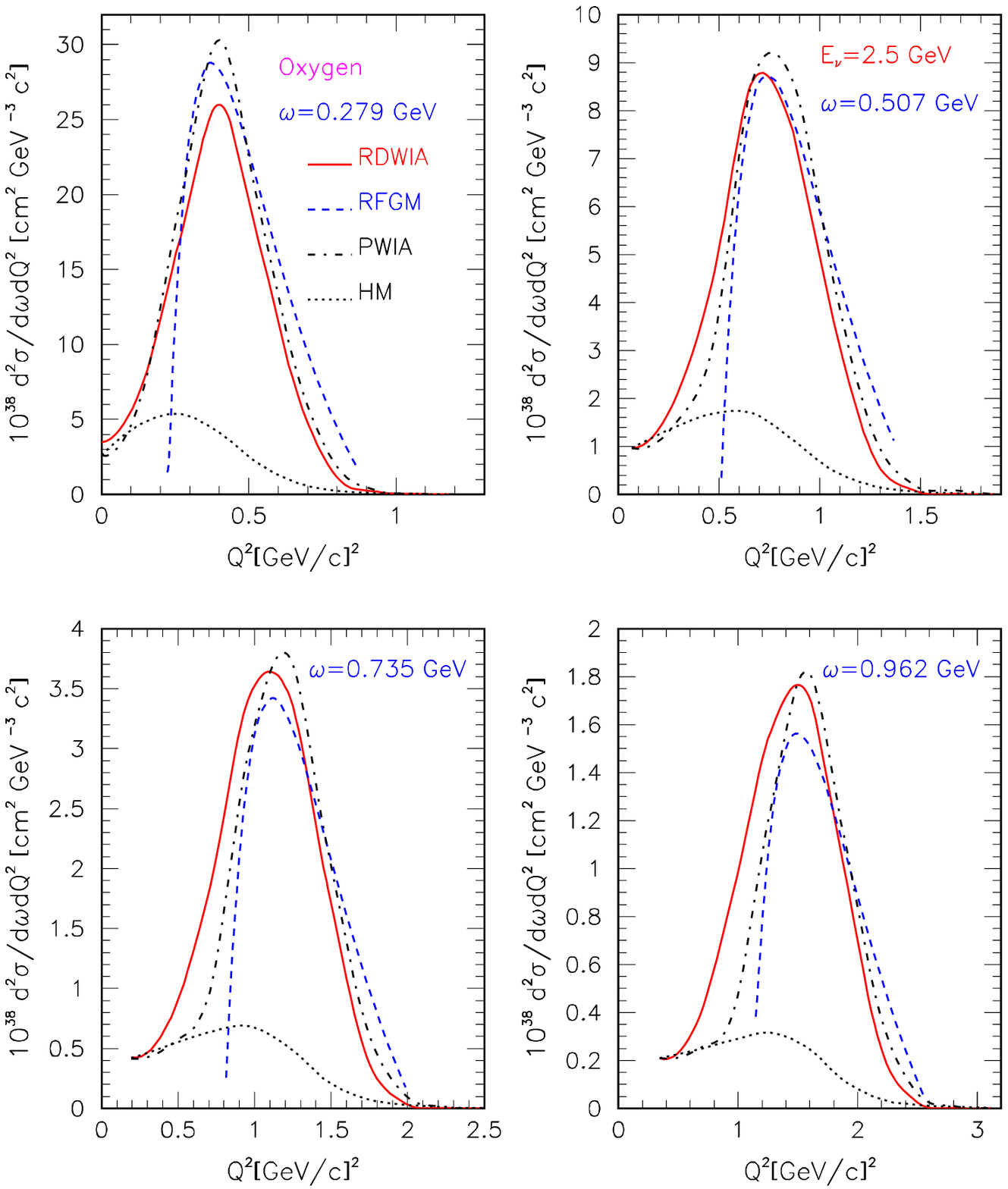}  
\end{center}
\caption{(Color online) Same as Fig.1, but for the neutrino energy 
$\varepsilon_{\nu}$=2.5 GeV and for four values of energy transfer:
$\omega$=0.279, 0.507, 0.735 and 0.962 GeV.
}
\end{figure*}
\begin{figure*}
  \begin{center}
    \includegraphics[height=16cm,width=16cm]{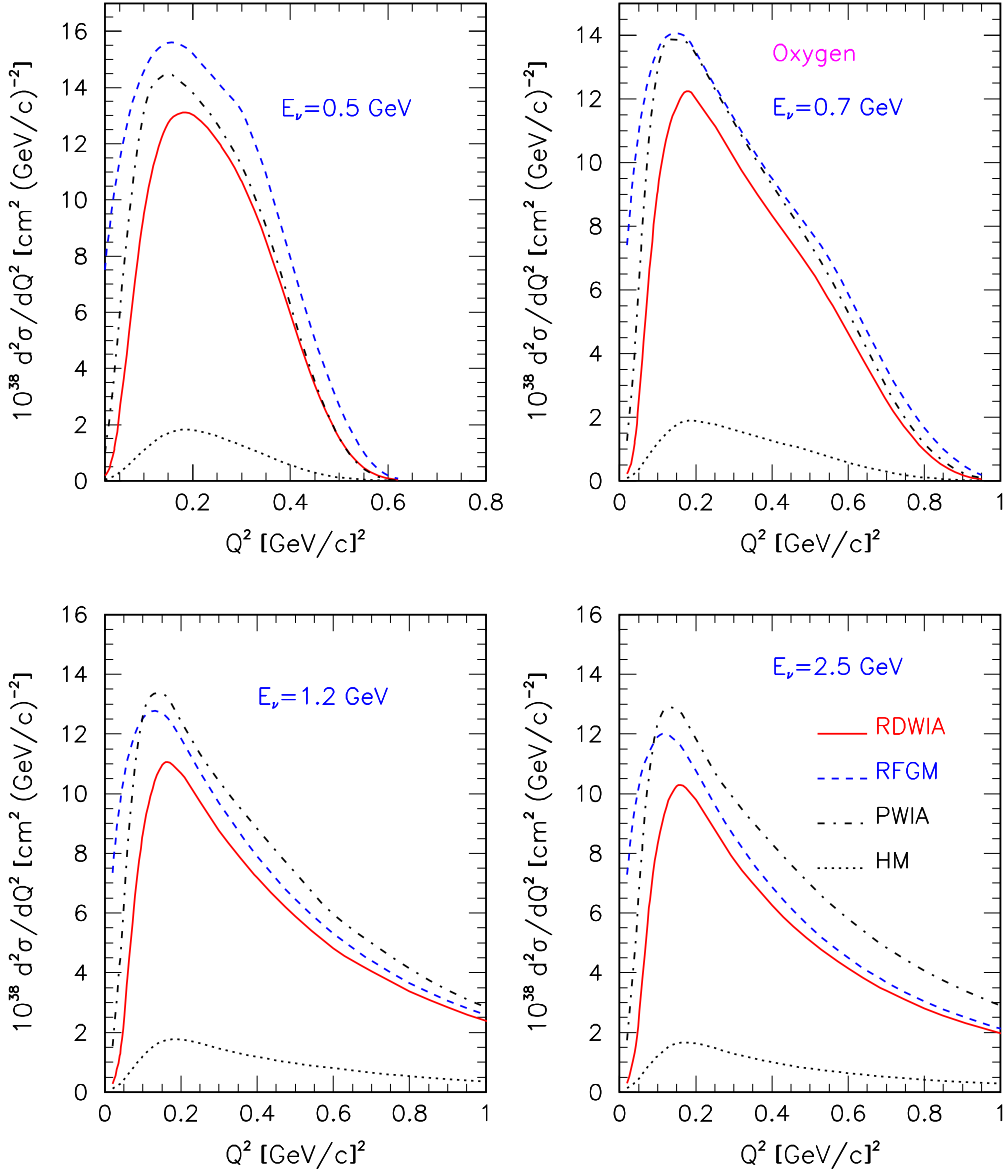}
  \end{center}
\caption{(Color online) Inclusive cross section versus the four-momentum 
transfer $Q^2$ for neutrino scattering off $^{16}$O and for four values of 
incoming neutrino energy: $\varepsilon_{\nu}$=0.5, 0.7, 1.2 and 2.5 GeV. 
The solid line is the RDWIA calculation, whereas the dashed and dash-dotted 
lines are, respectively, the RFGM and PWIA calculations. The dotted line is 
the high-momentum component contribution to the inclusive cross section.
}
\end{figure*}
\begin{figure*}
  \begin{center}
    \includegraphics[height=40pc,width=40pc]{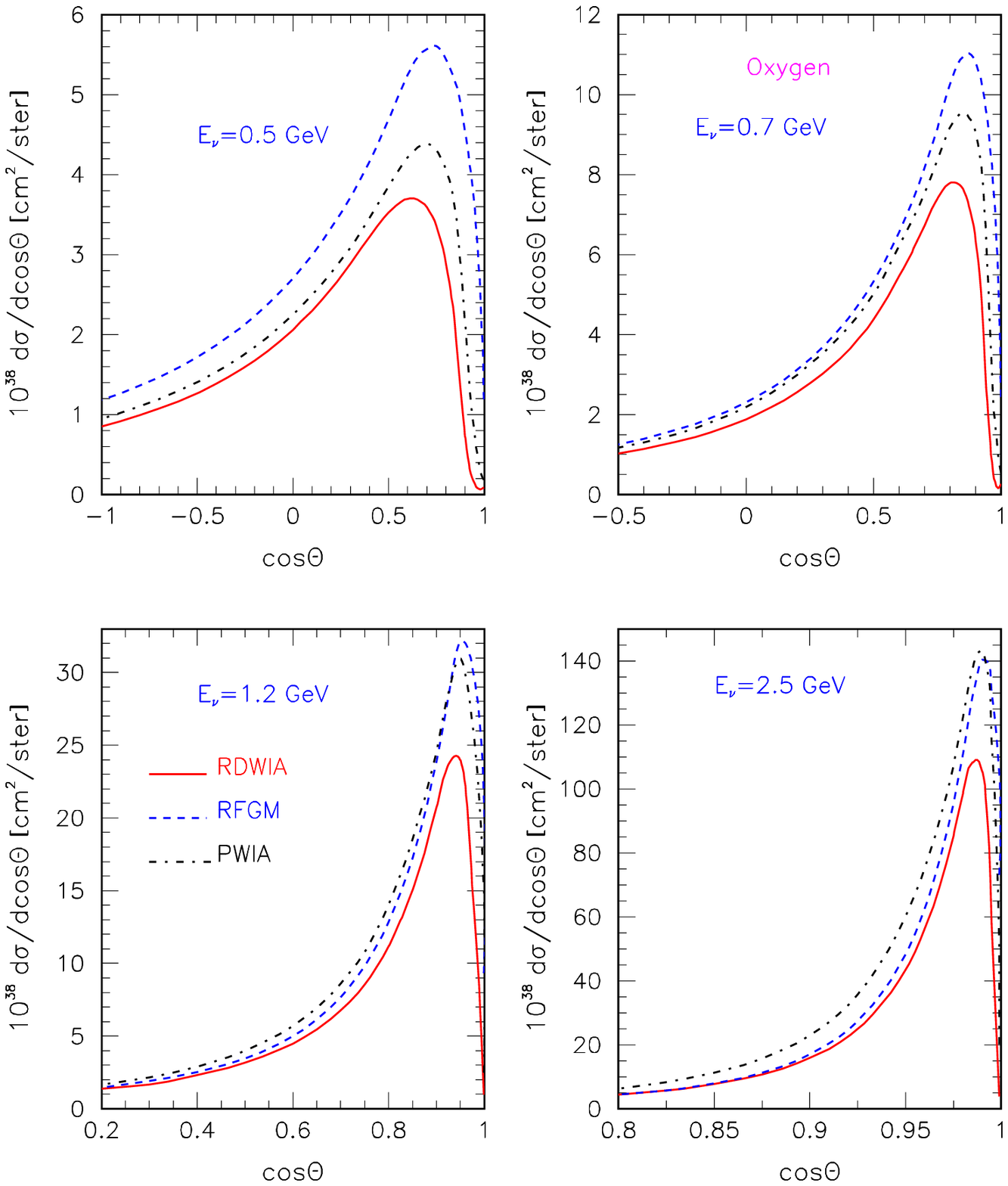}
  \end{center}
\caption{(Color online) Inclusive cross section versus the muon scattering
angle for four values of incoming neutrino energy: 
$\varepsilon_{\nu}$=0.5, 0.7, 1.2 and 2.5 GeV. The solid line is the RDWIA 
calculation, whereas the dashed and dash-dotted lines are, respectively, the 
RFGM and PWIA calculations.
}
\end{figure*} 

\section{Results}

The resulting fluxes of neutrino are predicted with the mean energy of $\sim$
0.7 GeV in the MiniBooNE and T2K experiments, and $\sim$ 2.5 GeV at 
the MINOS and MINERvA detectors. We calculated the differential inclusive cross
sections $d^3\sigma/d\omega dQ^2$ and $d^2\sigma/d\omega d\cos\theta$ of 
CCQE $\nu_{\mu}$ scattering off $^{16}$O for these energies using the LEA 
code, which was adopted for neutrino interaction~\cite{But2}. In Fig.1 
$d^3\sigma/d\omega dQ^2$ cross sections, calculated within the RDWIA, PWIA, 
and RFGM, are shown for the neutrino energy $\varepsilon_{\nu}$=0.7 GeV, 
and in Fig.2 for $\varepsilon_{\nu}$=2.5 GeV as a function of $Q^2$. Also 
shown in Figures 1 and 2 show the high-momentum component contributions of the 
nucleon momentum distribution in the target.

At energy $\varepsilon_{\nu}$=0.7 GeV the RDWIA cross sections, in the maximum,
are lower, than the PWIA and RFGM results, and this difference decreases with
increasing energy transfer. At energy 2.5 GeV, in the range of 
$\omega \geq$ 0.5 GeV, around the peak, the RFGM results are lower, than the 
RDWIA ones and fall down rapidly as $Q^2$ decreases. This trend is 
characteristic of the nucleon momentum distribution and the Pauli blocking 
effect as calculated in the Fermi gas 
model. On the contrary, the $Q^2$-dependence of the RDWIA and PWIA 
cross sections at low $Q^2$ is softer due to the HM-component contribution, 
which becomes dominant at $Q^2 <$ 0.1 (GeV/c)$^2$.    

Generally, theoretical uncertainties of the correlated NN-pairs contribution 
to the inclusive cross sections are higher as compared to the shell-nucleons 
contribution. The electron-nucleus scattering data [14, 32, 33] show that more 
complicated configurations, than a simple hard interaction between two 
nucleons, are involved in this case. Moreover, the off-shell ambiguities will 
be important for the high-momentum component, and one might expect the details 
of the off-shell extrapolation to become critical [34].

The inclusive cross sections for energies $\varepsilon_{\nu}$=0.5, 0.7, 1.2,
and 2.5 GeV are presented in Fig.3, which shows $d^2\sigma/Q^2$ as a function
of $Q^2$. Here the results, obtained in the RDWIA, are compared with cross
sections calculated in the PWIA and RFGM. The contributions of the
NN-correlations are shown as well. The cross sections, calculated in the 
Fermi gas model, are higher, than those obtained within the RDWIA, and this 
difference increases with decreasing $Q^2$. At $Q^2$=0.1 (GeV/c)$^2$ this 
discrepancy equals 54\% for $\varepsilon_{\nu}$=0.5 GeV and 43\% for 
$\varepsilon_{\nu}$=2.5 GeV. In the region around the maximum 
$Q^2$=0.2 (GeV/c)$^2$ the difference is about $\sim$18\% for 
$\varepsilon_{\nu}$=0.5 GeV and $\sim$11\% for $\varepsilon_{\nu}$=2.5 GeV. 
At $Q^2$=0.05 (GeV/c)$^2$ the contribution of the HM-component increases with 
energy from $\sim$ 15\% up to 23\% in the energy range of 0.5 $\div$ 2.5 GeV. 

Figure 4 shows the inclusive cross sections $d\sigma/d\cos\theta$ calculated
in the RDWIA, PWIA, and RFGM approaches for energies $\varepsilon_{\nu}$=0.5,
0.7, 1.2, and 2.5 GeV. They are displayed as a function of $\cos\theta$. It is
clear that in the region 0.8$<\cos\theta<$1 the values of the RFGM cross 
sections are higher, than those obtained within the RDWIA, and this difference 
decreases with neutrino energy. For energy $\varepsilon_{\nu}$=0.5 GeV 
($\varepsilon_{\nu}$=2.5 GeV) this discrepancy is 
about 25 times ($\sim$11\%) at $\cos\theta$=0.95 and $\sim$89\% ($\sim$2\%)
at $\cos\theta$=0.8. We note that measured $Q^2$ and
$\cos\theta$-distributions of the CCQE events [2, 11, 12] show similar 
features as compared to the RFGM prediction.

Figures 5 and 6 show $\cos\theta^m_{pq}$ as a function of $Q^2$ calculated
within the RDWIA and Fermi gas model kinematics for energies 
$\varepsilon_{\nu}$=0.5 and 2.5 GeV. The outgoing proton carries the kinematic
energy, that is approximately $\omega$. So far as  $\omega$ is low, 
the problem consists in identifying the events with very soft recoil proton; 
for high $\omega$ this
proton has high energy and may interact in the detector, making particle
identification and track reconstruction more challenging. In these figures we
show the contours of the phase volume in the ($\cos\theta_{pq}, Q^2$) 
coordinates for 0.25$\leq \omega \leq$1 GeV. Apparently, in the RDWIA 
kinematics this volume is larger, than in the RFGM. On the other hand, the 
difference decreases with $\omega$ and neutrino energy. Thus, systematic 
errors for the efficiency and purity of the two-track events selection are 
nuclear-model dependent.

We have studied the accuracy of the neutrino energy reconstruction with 
neglecting the systematics related to the event selection and resolution, 
i.e. with no detector effects or background. The study was performed with the 
values of cuts $(k_f)_{cut}$=0.2 GeV/c and $(\cos\theta)_{cut}$=0.

In Fig.7 the uncertainties of the energy reconstruction using Eq.(\ref{Rest})
within the RDWIA and RFGM approaches are presented as functions of neutrino
energy. The top panel shows the bias $\Delta=(\varepsilon_i -
\varepsilon_r)/\varepsilon_i$, the middle panel shows the variance
$\sigma/\varepsilon_i$ (the energy resolution), and the efficiency of the 
one-track events detection is displayed in the bottom panel. It is clear 
that in the case of the Fermi gas model Eq.(\ref{Rest}) systematically 
underestimates the neutrino energy and $\Delta$ decreases as the energy 
increases from $-$4.7\% for 
$\varepsilon_{\nu}$=0.3 GeV up to $-$0.7\% for $\varepsilon_{\nu}$=2.5 GeV. 
The variance $\sigma/\varepsilon_i$ (efficiency) increases with energy from 
$\sim$5.4\% ($\sim$71\%) up to $\sim$12\% ($\sim$99\%) over the range of
energy from 0.3 to 2.5 GeV. 
\begin{figure*}
  \begin{center}
    \includegraphics[height=16cm,width=16cm]{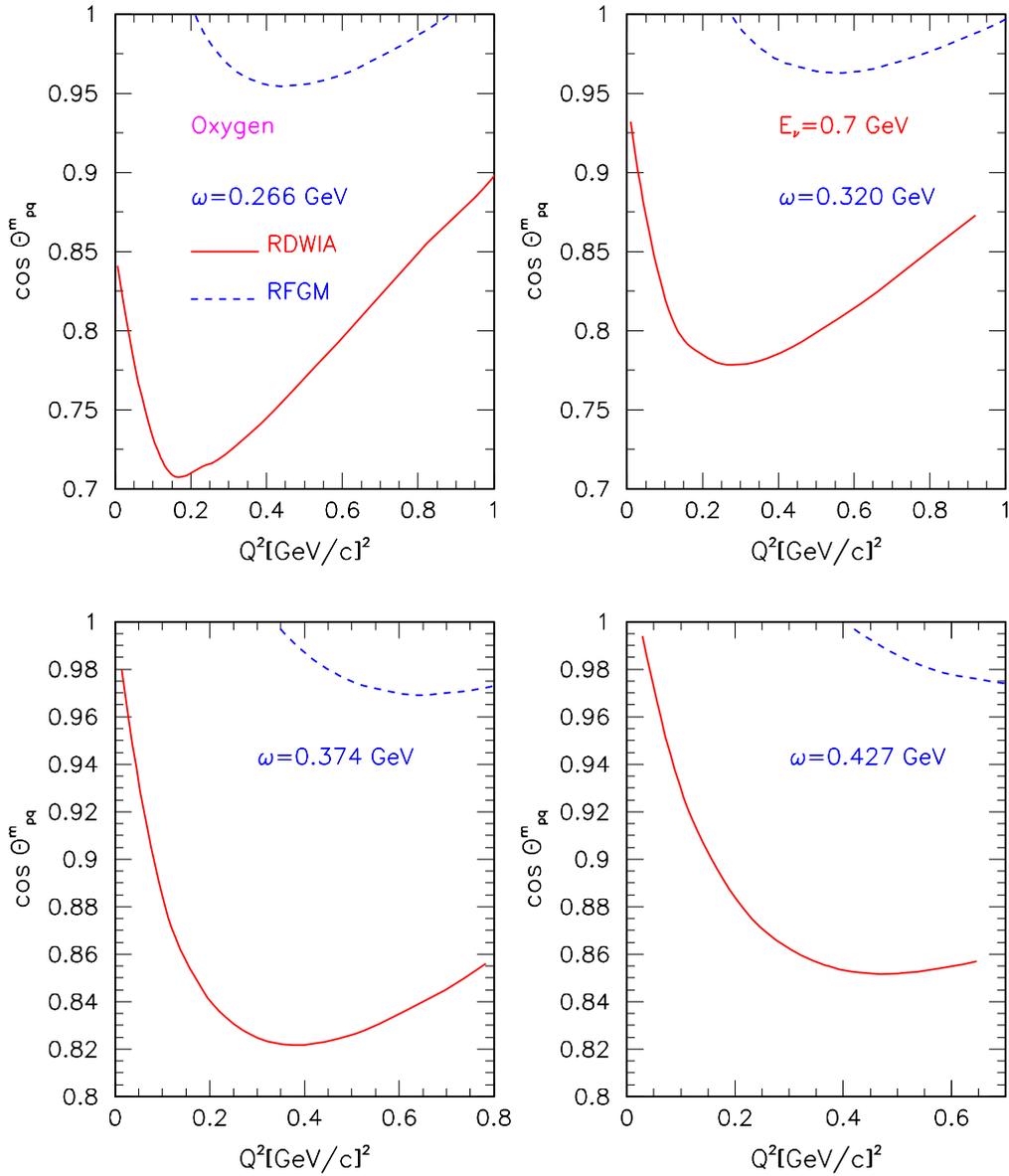}
  \end{center}
  \caption{(Color online) Contours of the phase volume in the
    ($\cos\theta_{pq}, Q^2$) coordinates for neutrino scattering off $^{16}$O 
with energy $\varepsilon_{\nu}$=0.7 GeV and for four values of energy 
transfer: $\omega$=0.288, 0.320, 0.374 and 0.427 GeV. The solid line is the 
RDWIA calculation, whereas the dashed line is the RFGM calculation. 
}
\end{figure*}
\begin{figure*}
  \begin{center}
    \includegraphics[height=16cm,width=16cm]{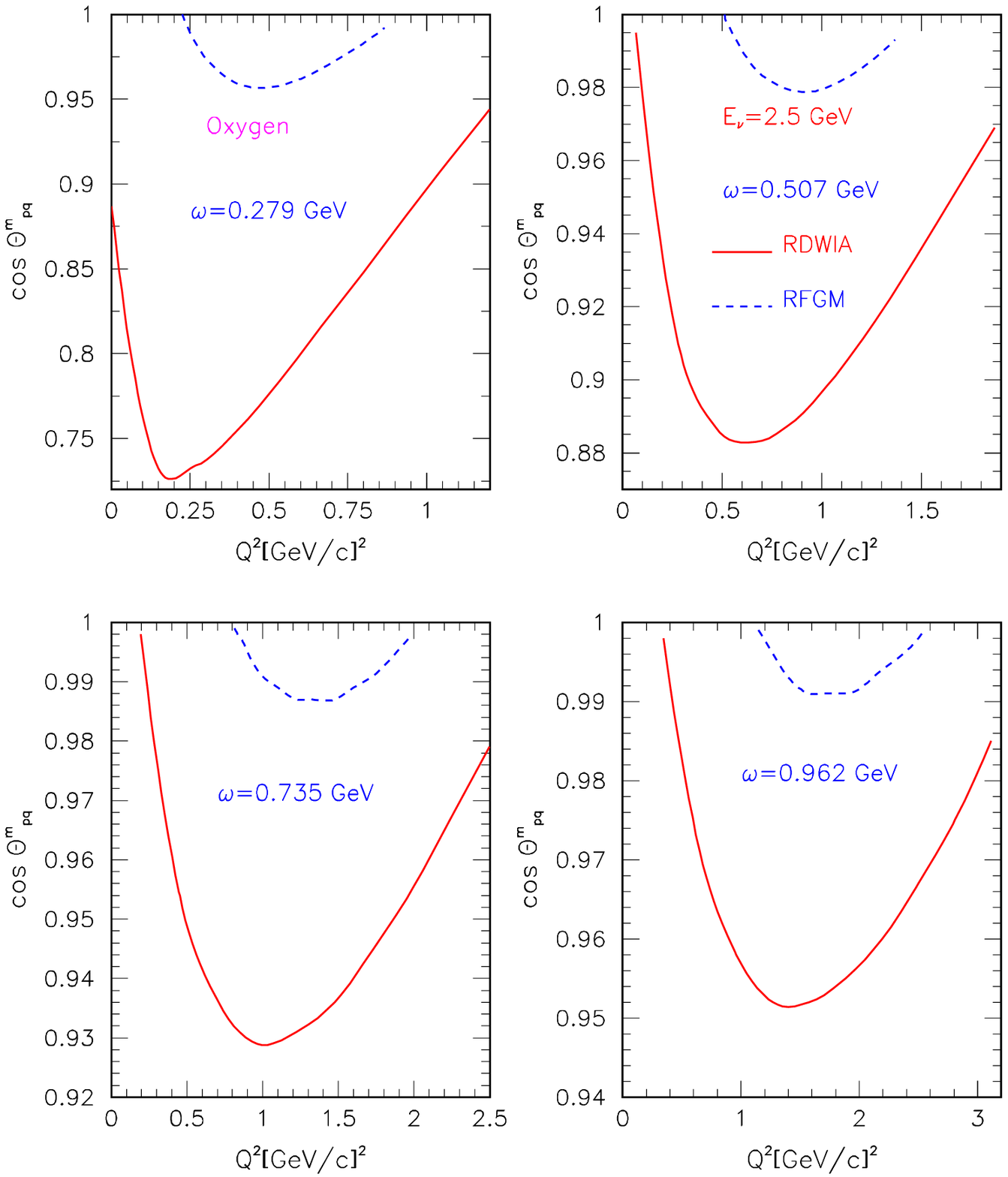}
  \end{center}
  \caption{(Color online) Same as Fig.5, but for neutrino energy 
$\varepsilon_{\nu}$=2.5 GeV and for four values of energy transfer:
$\omega$=0.279, 0.507, 0.735 and 0.962 GeV.
}
\end{figure*}
\begin{figure}
  \begin{center}
    \includegraphics[height=16cm,width=15cm]{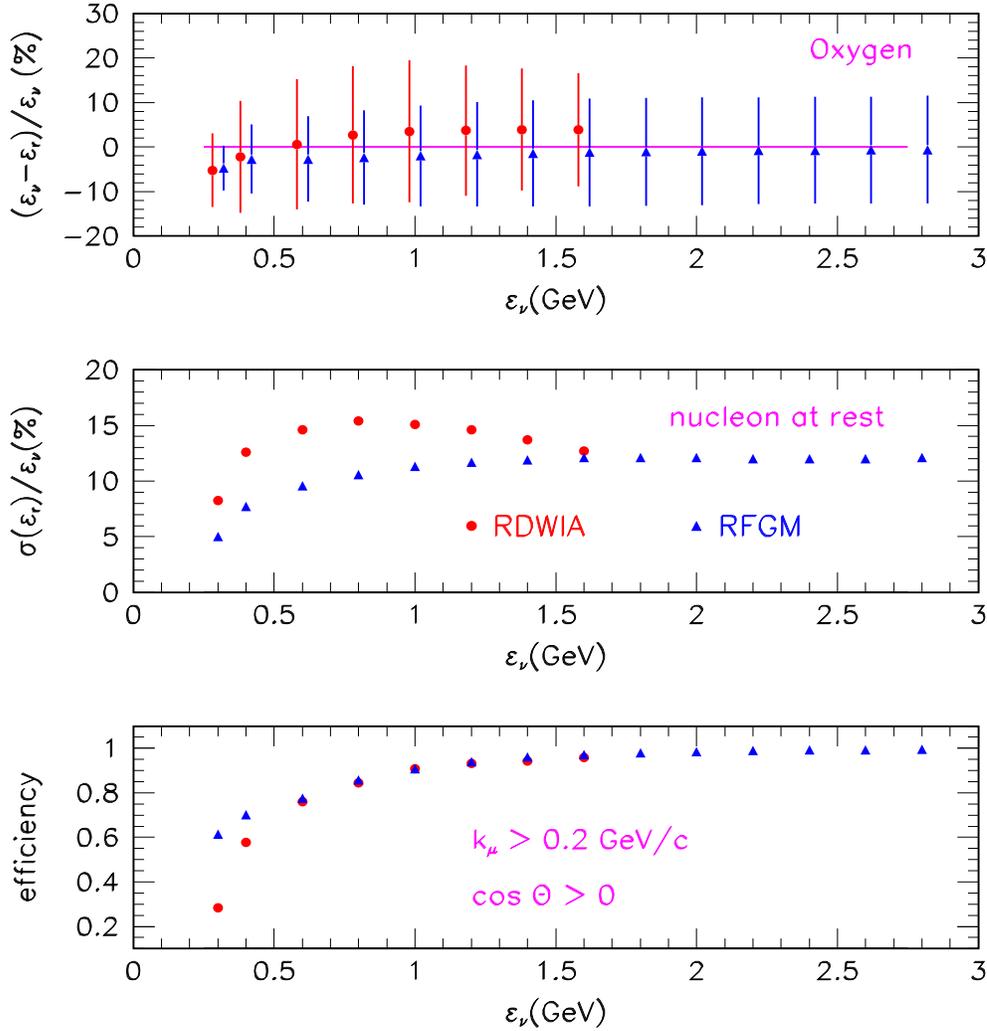}
  \end{center}
  \caption{(Color online) Bias (top panel), variance (middle panel) of the
    reconstructed neutrino energy, and the efficiency (bottom panel) of the 
    one-track events detection with $k_f\geq$0.2 (GeV/c) and 
    $\cos\theta\geq$0 as functions of neutrino energy. The neutrino energy 
    reconstruction was formed assuming the target nucleon to be at rest inside
    the nucleus. The vertical bars show $\sigma[(\varepsilon_i-
    \varepsilon_r)/\varepsilon_i]$. As displayed in the key, the biases, 
    variances and efficiencies were calculated in the RDWIA and RFGM.  
}
\end{figure}
\begin{figure}[t]
  \begin{center}
   \includegraphics[height=16cm, width=15cm ]{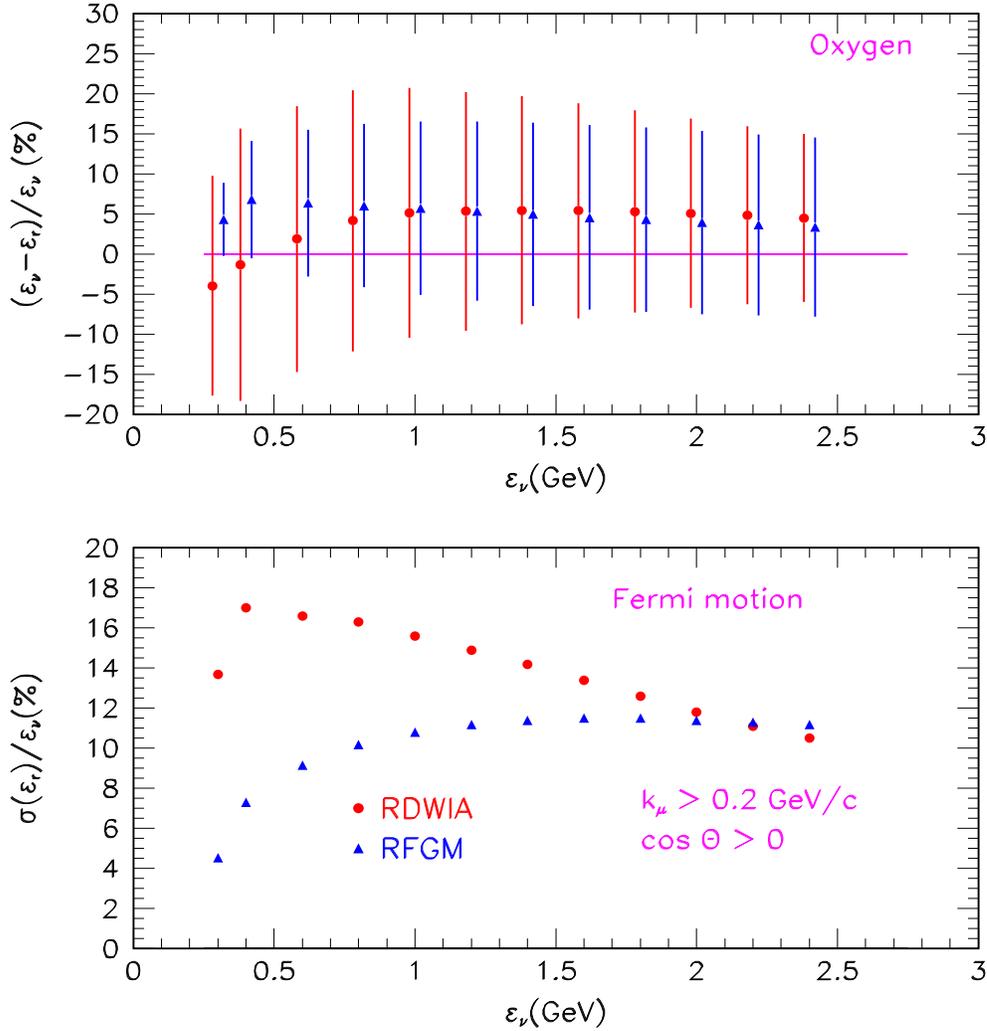}
  \end{center}
  \caption{(Color online) Bias (top panel) and variance (bottom panel) of the 
reconstructed neutrino energy as functions of neutrino energy. The energy 
reconstruction was formed taking into account the nucleon momentum
distribution in the target. The vertical bars are the same as in Fig.7. As 
displayed in the key, the biases and variances were calculated in the RDWIA 
and RFGM. 
}
\end{figure}
\begin{figure}
  \begin{center}
    \includegraphics[height=16cm,width=15cm]{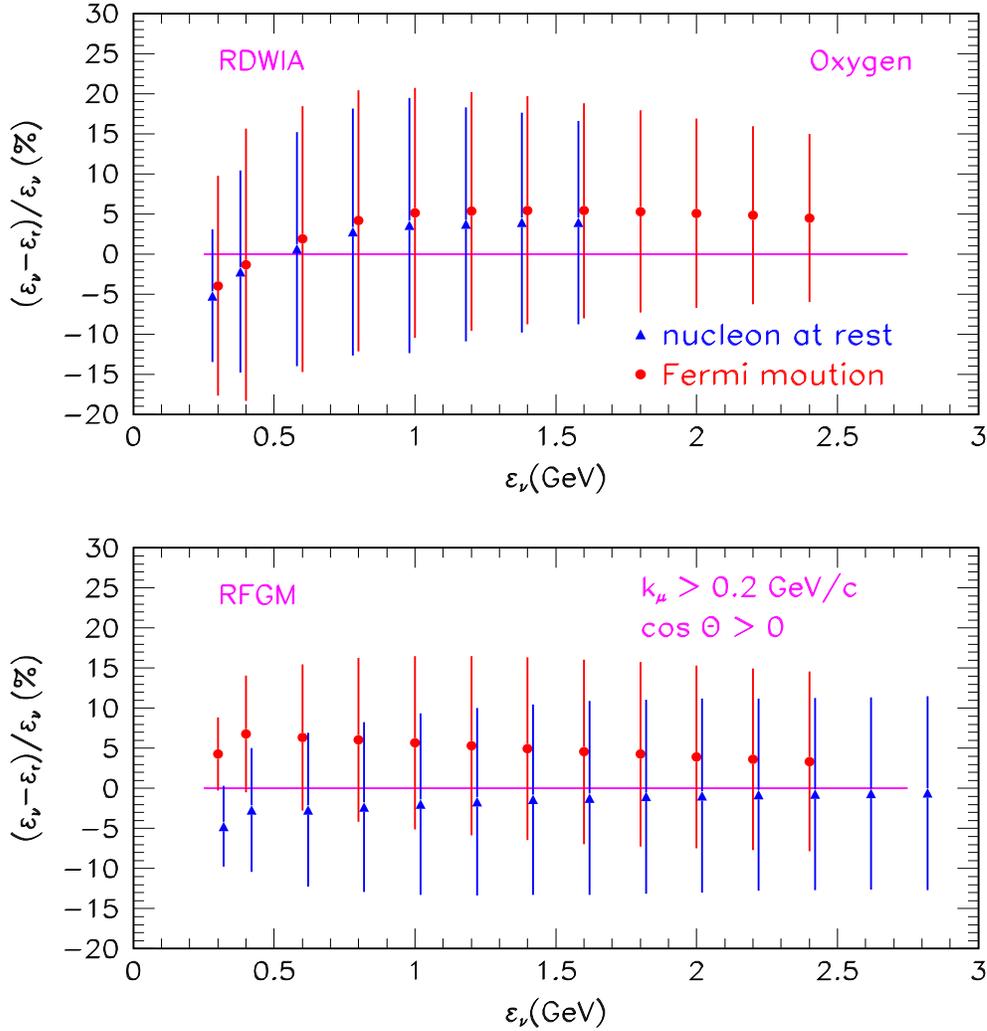}
  \end{center}
  \caption{(Color online) Biases calculated in the RDWIA (top panel) and RFGM 
(bottom panel) as functions of neutrino energy. The vertical bars are the 
same as in Fig.7. As displayed in the key, the energy reconstructions were 
formed with and without the nucleon momentum distribution.  
}
\end{figure}
\begin{figure}
  \begin{center}
    \includegraphics[height=16cm,width=15cm]{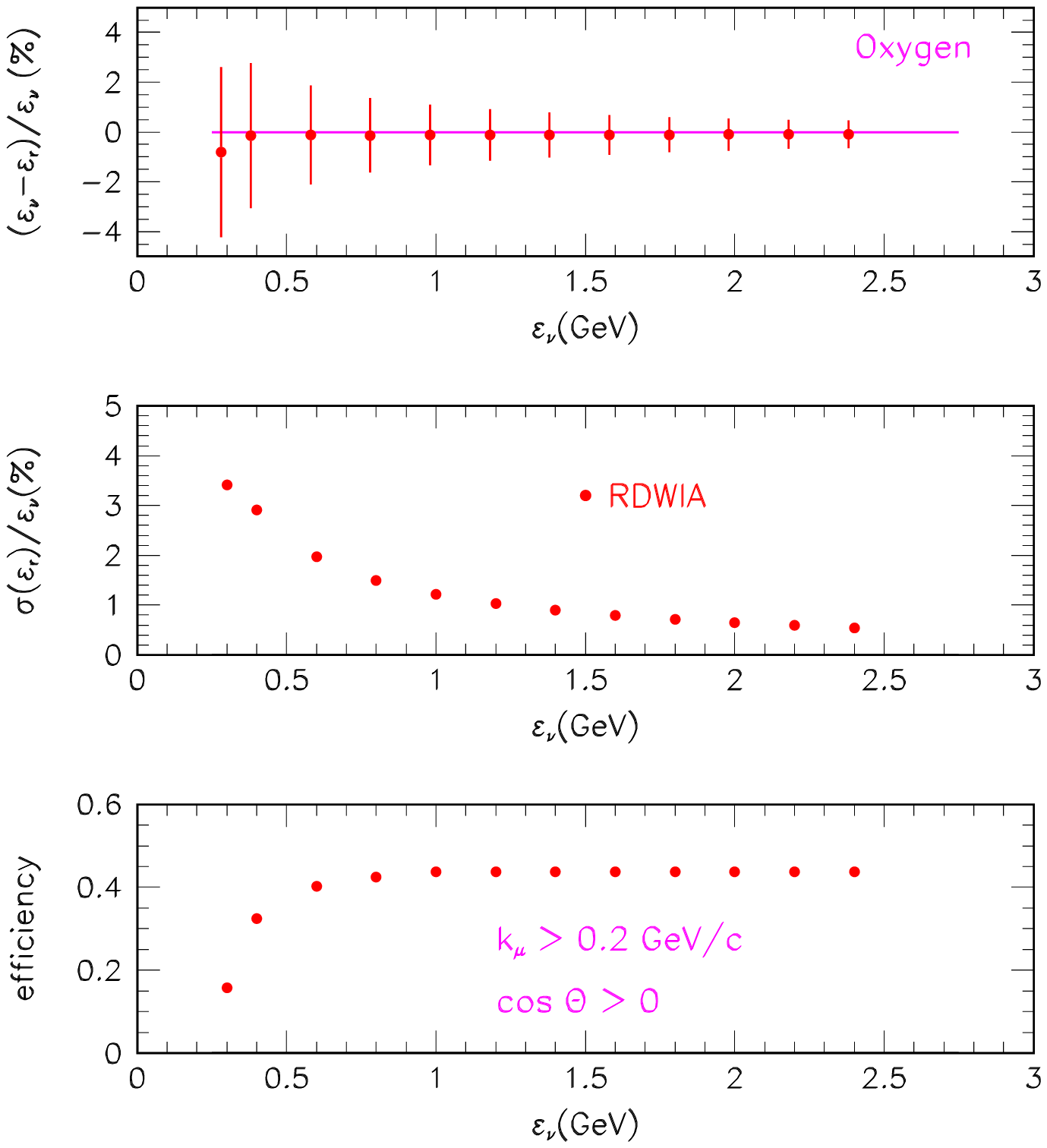}
  \end{center}
  \caption{(Color online) Bias (top panel), variance (middle panel) of the
    reconstructed neutrino energy, and the efficiency (bottom panel) of the 
    two-track events detection with $k_f\geq$0.2 (GeV/c) and 
    $\cos\theta\geq$0. The vertical bars are the same as in Fig.7. The bias, 
    variance and efficiencies were calculated within the RDWIA. 
}
\end{figure}

In the RDWIA approach $\Delta\approx-$5.2\% for $\varepsilon_{\nu}$=0.3 GeV
and $\Delta\approx$3.9\% for $\varepsilon_{\nu}$=1.6 GeV. At energies 
$\varepsilon_{\nu} > $1.6 GeV and at fixed values of cuts the
denominator in Eq.(\ref{Rest}) can be negative in the detection volume of 
the ($k_f$, $\cos\theta$) phase space, which is determined by 
Eqs.(\ref{neg1}) and (\ref{neg2}). We note that the size of this volume 
decreases as $(\cos\theta)_{cut}$ increases. The energy resolution is about 
8.3\% for $\varepsilon_{\nu}$=0.3 GeV and 
$\sim$12.7\% for $\varepsilon_{\nu}$=1.6 GeV, and the maximum of 15.4\% is 
located around $\varepsilon_{\nu}$=0.8 GeV. The efficiency rapidly increases 
with energy from 28\% for $\varepsilon_{\nu}$=0.3 GeV up to 96\% for 
$\varepsilon_{\nu}$=1.6 GeV. So, the values of the bias and energy resolution, 
obtained within the RDWIA, are higher, than those obtained in the Fermi gas 
model.    

The accuracy of the mean energy method, which takes into account the nucleon 
momentum distribution in the target, is shown in Fig.8. We assume that the 
maximum neutrino energy in the neutrino beam (Eq.(\ref{z_l})) is $E_{max}$=10 
GeV. The top and bottom panels show the biases and energy resolutions 
calculated within the RDWIA and RFGM. In the case of the RFGM the mean energy 
method systematically overestimates the neutrino energy; $\Delta$=4.3\% for 
$\varepsilon_{\nu}$=0.3 GeV, $\Delta\approx$6\% for 
$\varepsilon_{\nu}$=0.8 GeV and decreases down to 3\% for 
$\varepsilon_{\nu}$=2.5 GeV. The energy resolution increases with energy from 
4.6\% up to 11\% in this energy range. In the RDWIA approach 
$\Delta\approx-$4\%, $(\sigma/\varepsilon_i)\approx$14\% for 
$\varepsilon_{\nu}$=0.3 GeV, and 
$\Delta\approx$4.5\%, $(\sigma/\varepsilon_i)\approx$10.5\% for 
$\varepsilon_{\nu}$=2.5 GeV. It should be noted here that bias may depend on
the value of $E_{max}$. 

The effect of the nucleon momentum distribution in the target is shown in
Fig.9. The biases, calculated within the RDWIA (top panel) and RFGM (bottom
panel) using Eq.(\ref{Rest}) ( $\Delta_{fr}$ ) and the mean energy method 
( $\Delta_{me}$ ), are presented as functions of neutrino energy. In the RDWIA
approach the $\Delta_{fr}$ and $\Delta_{me}$ show similar behavior with
neutrino energy, and the nucleon Fermi motion effect leads to increasing 
the bias by about 1.2\%. In the Fermi gas model with this effect 
$\varepsilon_r$ is overestimated, 
and $\Delta_{fr}(\Delta_{me})=-$4.7\%(4.3\%) for energy 0.3 GeV and 
$\Delta_{fr}(\Delta_{me})=-$0.7\%(3.4\%) for $\varepsilon_{\nu}$=2.5 GeV.

Apparently, the accuracy of the kinematic reconstruction of neutrino energy for
one-track events depends on the nuclear models of QE neutrino CC interaction 
with nuclei and on the neutrino energy reconstruction methods. In the K2K 
and MiniBooNE experiments Eq.(\ref{Rest}) is applied for the energy 
reconstruction. The bias $\Delta_{FG}$ and energy resolution 
$\delta_{FG}=(\sigma/\varepsilon_i)_{FG}$ were calculated using MC simulation 
based on the Fermi gas model. We can estimate
the systematic uncertainties of this approach by comparing $\Delta_{FG}$
and $\delta_{FG}$ with $\Delta_R$ and $\delta_R$ evaluated in the RWDIA 
approach using the mean energy method. It is clear that uncertainties depend 
on neutrino energy: $\Delta_{FG}(\Delta_R)\approx-$4.7\%($-$4\%) and
$\delta_{FG}(\delta_R)\approx$5.4\%(13.7\%) for $\varepsilon_{\nu}$=0.3 GeV; 
$\Delta_{FG}(\Delta_R)\approx-$2.3\%(4.1\%) and
$\delta_{FG}(\delta_R)\approx$10.6\%(16.3\%) for $\varepsilon_{\nu}$=0.8 GeV; 
$\Delta_{FG}(\Delta_R)\approx-$0.7\%(4.5\%) and
$\delta_{FG}(\delta_R)\approx$12\%(11.5\%) for $\varepsilon_{\nu}$=2.5
GeV. So, the bias uncertainty increases with energy from
$(\Delta_R-\Delta_{FG})\approx$0.7\% for $\varepsilon_{\nu}$=0.3 GeV up to
5.2\% for $\varepsilon_{\nu}$=2.5 GeV, and the energy resolution uncertainty
decreases with increasing energy from $\delta_R-\delta_{FG}\approx$8.3\% 
down to 0.5\% in this energy range. We note that these estimations may depend 
on the values of $(k_f)_{cut}$, $(\cos\theta)_{cut}$ and $E_{max}$.           

In Fig.10 the accuracy of the energy reconstruction for the two-track events,
calculated using Eq.(\ref{calor}) within the RDWIA, is shown as a function of
neutrino energy. The top panel shows the bias, the middle panel shows the
variance, and the bottom panel shows the efficiency of the two-track events
detection with cuts $k_f\geq$0.2 GeV/c and $\cos\theta\geq$0 for muon tracks
and without any cuts for the proton tracks. At energy $\varepsilon_{\nu}>$0.3
GeV the bias is $\Delta\approx-$0.1\% and does not depend on the neutrino 
energy. The energy resolution decreases as the energy increases from 3.4\% for 
$\varepsilon_{\nu}$=0.3 GeV up to 0.5\% for $\varepsilon_{\nu}$=2.5 GeV,
and the efficiency rapidly increases with energy from $\sim$16\% up to 
$\sim$44\% in this energy range. These estimations show that the accuracy of 
the calorimetric method can be higher, than the kinematic one and does not 
depend on the model of CC neutrino QE interaction with nuclei and on the 
nucleon momentum distribution in the target. The challenge is identifying the 
proton track and reconstructing its kinetic energy with reliable accuracy at 
the low threshold energy for proton
detection. 

\section{Conclusions}

In this paper, we study the quasi-elastic neutrino charged-current 
scattering on the oxygen target in various approximations (PWIA, RDWIA,
RFGM) making particular emphasis on the nuclear-model dependence of the
results.  In the RDWIA, the LEA program, adapted to neutrino interactions, 
was used for calculating the differential cross sections with the effect of 
NN-correlations in the target ground state.

The inclusive $d^2\sigma/dQ^2$ and $d\sigma/d\cos\theta$ cross sections, 
calculated within the RDWIA, and the measured $Q^2$, 
$\cos\theta$-distributions of 
CCQE events exhibit similar feature as compared to the Fermi gas model. The 
magnitude of inclusive cross sections 
$d^2\sigma/dQ^2$ and $d\sigma/d\cos\theta$ 
is lower in the RDWIA calculations, than that of the Fermi gas model, and in 
the region around the maximum $Q^2$=0.2 (GeV/c)$^2$ the difference is about 
18\% for $\varepsilon_{\nu}$=0.5 GeV and 11\% for $\varepsilon_{\nu}$=2.5
GeV. The contribution of the HM-component at $Q^2$=0.05 (GeV/c)$^2$ increases 
with neutrino energy from 15\% up to 23\% in this energy range. Note that the 
measured $Q^2$ and $\cos\theta$-distributions of CCQE events are also lower, 
than the RFGM prediction at low $Q^2$.

We have shown that the efficiency and purity of the CCQE two-track events
selection are nuclear-model dependent, and the difference decreases with 
increasing energy transfer and neutrino energy.

We have studied the nuclear-model dependence of the energy reconstruction 
accuracy, neglecting the systematics related to event selection and 
resolution. We found that the accuracy of the kinematic reconstruction for 
one-track events depends on the nuclear model of CCQE neutrino interaction 
and on the neutrino energy reconstruction method. The uncertainties in the 
reconstructed energy bias increase in the energy range of 0.3 $\div$ 2.5 GeV 
from $\sim$0.7\% up to 5.4\%, and the energy resolution ambiguities decrease 
from 8.3\% down to 0.5\%  with 
increasing energy. In the case of two-track events the accuracy may be 
higher and does not depend on the nuclear models of CCQE neutrino-nucleus 
interaction. 

We conclude that the use of RDWIA in the Monte Carlo simulation of neutrino
detector and the data analysis would allow one to reduce the systematic 
uncertainty in neutrino oscillation parameters.

\section*{Acknowledgments}

The author greatly acknowledges S. Kulagin, S. Mikheyev, J. Morfin, G. Zeller, 
and R. Gran for fruitful discussions at different stages of this work.  

\appendix
\section{Equation for neutrino energy}
\label{A}

In Eq.(\ref{Equ}) 
$$
A\varepsilon^2_r - B\varepsilon_r + C= 0 .                 
$$
the coefficients A, B, and C are defined as follows:
\begin{subequations}\label{ABC}
\begin{align}
A & = a^2 - \p^2_m z^2,
\\
B & = ab - 2\p^2_m z^2k_f\cos\theta,                 
\\
C & = b^2/4 - \p^2_m z^2 k^2_f,                 
\end{align}
\end{subequations}
where 
\begin{subequations}\label{abz}
\begin{align}
a & = \varepsilon_{ef} - k_f\cos\theta,
\\
b & = \varepsilon^2_{ef} - (\vert\p_m\vert^2+m^2) - k^2_f,            
\\
z &=\cos\tau=\p_m\cdot\q/|\p_m\cdot\q|.                 
\end{align}
\end{subequations}
In the Fermi gas model
\begin{equation}\label{ef:fg}
\varepsilon_{ef}=\varepsilon_f - \sqrt{\p^2_m+m^2}+\epsilon_b,      
\end{equation}
in the RDWIA model, for shell-nucleons
\begin{equation}\label{ef:shell}
\varepsilon_{ef}=\varepsilon_f - m_A + 
[(m_A-m+\varepsilon_m)^2+\p^2_m]^{1/2},                              
\end{equation}
and for nucleons in the correlated NN-pair
\begin{equation}\label{ef:nn}
\varepsilon_{ef}=\varepsilon_f - \varepsilon_N.                     
\end{equation}

\section{Moments of reconstructed neutrino energy}
\label{B}

In the Fermi gas model with the p.d.f.
\begin{equation}
S(\p_m,\varepsilon_m) =\frac{3}{4\pi p^3_F}\delta(\varepsilon_m-\epsilon_b)  
\end{equation}                                                      
Eq.(\ref{mom1}) takes the form 
\begin{equation}\label{FG:mom}
\langle\varepsilon^n_r(k_f,\cos\theta)\rangle =
\frac{3}{4\pi p^3_F}
\int_{p_{min}}^{p_{max}}p^2dp \int_{z_{min}}^{z_{max}}               
[\varepsilon_r(k_f,\cos\theta,p,z)]^n dz,
\end{equation}
where $z=\cos\tau$ and $\varepsilon_r$ is given by Eqs.(\ref{Erec}), 
(\ref{ABC}), (\ref{abz}), and (\ref{ef:fg}). In the RDWIA the p.d.f. can be 
written as follows:
\begin{equation}
S(\p,\varepsilon) =\sum_{\alpha}v_{\alpha}S_{\alpha}(\p)
\delta[\varepsilon-(\varepsilon_m)_{\alpha}] +                        
v_{NN}S_{NN}(\p,\varepsilon)
\end{equation}
and we have
\begin{subequations}\label{RD:mom}
\begin{align}
\langle\varepsilon^n_r(k_f,\cos\theta)\rangle & = \sum_{\alpha}v_{\alpha}
\langle\varepsilon^n_{r,\alpha}(k_f,\cos\theta)\rangle + 
v_{NN}\langle\varepsilon_{r,NN}(k_f,\cos\theta)\rangle,
\\                                                                     
\langle\varepsilon^n_{r,\alpha}(k_f,\cos\theta)\rangle & =
\int_{p_{min}}^{p_{max}}p^2dp \int_{z_{min}}^{z_{max}}
S_{\alpha}(\p)[\varepsilon_{r,\alpha}(k_f,\cos\theta,p,z)]^n dz,
\\
\langle\varepsilon^n_{r,NN}(k_f,\cos\theta)\rangle & =
\int_{p_{min}}^{p_{max}}p^2dp 
\int_{\varepsilon_{min}}^{\varepsilon_{max}}d\varepsilon
\int_{z_{min}}^{z_{max}}
 S_{NN}(\p,\varepsilon)
[\varepsilon_{r,NN}(k_f,\cos\theta,p,\varepsilon,z)]^n dz,
\end{align}
\end{subequations}
where $S_{\alpha}$ and $S_{NN}$ are, respectively, the p.d.f. for the 
momentum and energy of nucleons on the shell $\alpha$ and in the correlated 
NN-pairs, $\varepsilon^n_{r,\alpha}$ and $\varepsilon^n_{r,NN}$ are given by 
Eqs.(\ref{Erec}), (\ref{ABC}), (\ref{abz}), (\ref{ef:shell}), (\ref{ef:nn}) 
and the sum is taken over occupied shells. The coefficients $v_{\alpha}$ and 
$v_{NN}$ are 
\begin{equation}\label{wei}
v_{\alpha,NN}  = \frac{1}{\sigma_{\alpha ,NN}}
\Big(\frac{d^2 \sigma}{dk_f d\cos\theta}\Big)_{\alpha,NN},           
\end{equation}
where
\begin{equation}
\sigma_{\alpha,NN} =\int\Big(\frac{d^2\sigma}
{dk_f d\cos\theta}\Big)_{\alpha,NN}dk_f d\cos\theta                  
\end{equation}
The integral is calculated with $(k_f)_{min}=(k_f)_{cut}$ and 
$(\cos\theta)_{min}=(\cos\theta)_{cut}$. 
Using Eqs.(\ref{mom}), (\ref{RD:mom}), we have
\begin{equation}
\langle\varepsilon^n_r(\varepsilon_i)\rangle =\sum_{\alpha}w_{\alpha}
\langle\varepsilon^n_r(\varepsilon_i)\rangle_{\alpha} +                
w_{NN}\langle\varepsilon^n_r(\varepsilon_i)\rangle_{NN},
\end{equation}
where $w_{\alpha,NN}=\sigma_{\alpha,NN}/(\sum_{\alpha}\sigma_{\alpha}+
\sigma_{NN})$.
In Eqs.(\ref{FG:mom}) and (\ref{RD:mom}) the limits of integration over $z$
are: $z_{min}$=-1 and $z_{max}$=min$\{1,z_l\}$. The value of $z_l$ is 
obtained from the requirement $\varepsilon_r(k_f,\cos\theta) \leq E_{max}$, 
where $E_{max}$ 
is the maximum neutrino energy in the neutrino beam. We note that this
constrain on $z_{max}$ leads to increasing bias $\Delta$ in the 
reconstructed energy. Then from Eqs.(\ref{p_x}), (\ref{eps_FG}), 
(\ref{eps_RD}), and (\ref{eps_NN}) with $\varepsilon_i=E_{max}$ it follows, 
that
\begin{equation}\label{z_l}
z_l=\frac{(\omega_{max}+\epsilon)^2-(\p^2_m+\q^2+m^2)}{2|\p_m||\q|},      
\end{equation}
where $\omega_{max}=E_{max}-\varepsilon_f$ and 
$\q^2=E^2_{max} + k^2_f - 2E_{max}k_f\cos\theta$.    
In the Fermi gas model $\epsilon=\sqrt{\p^2_m+m^2}-\epsilon_b$, in the RDWIA
for scattering off shell-nucleons $\epsilon=m_A-\sqrt{p^2_m+m^2_B}$, and
$\epsilon=\varepsilon_N$ for scattering off nucleons in the correlated NN-pair.

For the two-track events the moments of the $\varepsilon_r(k_f,\cos\theta)$ 
distribution can be written as 
\begin{equation}
\langle\varepsilon^n_r(\varepsilon_i)\rangle = \sum_{\alpha}w_{\alpha}\langle
\varepsilon^n_r(\varepsilon_i)\rangle_{\alpha},                        
\end{equation}
where
\begin{subequations}
\begin{align}
\langle\varepsilon^n_r(\varepsilon_i)\rangle & = \int dk_f \int d\cos\theta
\int_0^{2\pi}d\phi\int_{p_{min}}^{p_{max}} [\varepsilon_f + T_p + 
\langle\varepsilon_m\rangle]^n W_{\alpha}(k_f,\cos\theta,\phi, p_m)d p_m,
\\
W_{\alpha} & =\frac{1}{\sigma^{ex}_{\alpha}}\Big[\frac{d^5\sigma}
{dk_f d\cos\theta d\phi dp_m}\Big]_{\alpha}
\\                                                                    
\sigma^{ex}_{\alpha} & = \int dk_f \int d\cos\theta
\int_0^{2\pi}d\phi\int_{p_{min}}^{p_{max}} \Big[\frac{d^5\sigma}
{dk_f d\cos\theta d\phi dp_m}\Big]_{\alpha}d p_m,
\\
w_{\alpha} & =\sigma^{ex}_{\alpha}/\sum_{\alpha}\sigma^{ex}_{\alpha}
\end{align}
\end{subequations}
and $d^5\sigma/dk_f d\cos\theta d\phi dp_m$ is the QE neutrino CC scattering
exclusive cross section (\ref{cs:excl}) in terms of a missed momentum $p_m$.

\end{document}